\documentclass[iop,apj]{emulateapj}  

\usepackage[usenames,dvipsnames]{xcolor}

\newcommand{\kms}{km\,s$^{-1}$}

\renewcommand{\thefootnote}{\fnsymbol{footnote}}

\shorttitle{ZDI of II Peg in all four Stokes parameters}

\shortauthors{Ros\'{e}n et al.}

\begin{document}


\title{First Zeeman Doppler imaging of a cool star using all four Stokes parameters\footnotemark[*]}

\author{L.~Ros\'{e}n, O.~Kochukhov}
\affil{Department of Physics and Astronomy, Uppsala University, Box 516, Uppsala 751 20, Sweden}
\and
\author{G.~A.~Wade}
\affil{Department of Physics, Royal Military College of Canada, Box 17000, Station Forces, Kingston, Ontario K7K 7B4, Canada}

\begin{abstract}
Magnetic fields are ubiquitous in active cool stars but they are in general complex and weak. Current Zeeman Doppler imaging (ZDI) studies of cool star magnetic fields chiefly employ circular polarization observations because linear polarization is difficult to detect and requires a more sophisticated radiative transfer modeling to interpret. But it has been shown in previous theoretical studies, and in the observational analyses of magnetic Ap stars, that including linear polarization in the magnetic inversion process makes it possible to correctly recover many otherwise lost or misinterpreted magnetic features. We have obtained phase-resolved observations in all four Stokes parameters of the RS~CVn star II~Peg at two separate epochs. Here we present temperature and magnetic field maps reconstructed for this star using all four Stokes parameters. This is the very first such ZDI study of a cool active star. Our magnetic inversions reveal a highly structured magnetic field topology for both epochs. The strength of some surface features is doubled or even quadrupled when linear polarization is taken into account. The total magnetic energy of the reconstructed field map also becomes about 2.1--3.5 times higher. The overall complexity is also increased as the field energy is shifted towards higher harmonic modes when four Stokes parameters are used. As a consequence, the potential field extrapolation of the four Stokes parameter ZDI results indicates that magnetic field becomes weaker at a distance of several stellar radii due to a decrease of the large-scale field component.
\end{abstract}


\section{Introduction}
\label{intro}

All cool stars have a magnetic field since they can actively convert convective and rotational mechanical energy into electromagnetic energy through a dynamo mechanism. This magnetic field is not static but instead evolves over time. Fields are characterized by a continuum of scales, from highly localized to global, and a range of evolution timescales, from minutes to decades. Typical cool star magnetic fields are hence complex and changing structures, constantly evolving but at the same time relatively weak on global scales. This makes them challenging to investigate even though they are ubiquitous. However, stellar magnetic fields play an important role throughout a star's life, both for the star itself and for surrounding objects. It is therefore crucial to investigate them as carefully as possible.

\footnotetext{Based on observations obtained at the Canada-France-Hawaii Telescope (CFHT), which is operated by the National Research Council of Canada, the Institut National des Sciences de l'Univers of the Centre National de la Recherche Scientifique of France, and the University of Hawaii.}

\begin{table*}
\caption{Log of spectropolarimetric observations of II~Peg in all four Stokes parameters. The rotational phases in column three are calculated using the orbital ephemeris of $T_{\rm{maxRV}}=2448942.428 + 6.7242078E$. The numbers in column four represent the number of spectra acquired for each Stokes parameter. Uncertainties marked with * are mean values of the multiple measurements of that Stokes parameter.}
\label{tab1}
\centering
\begin{tabular}{ccccccr}
\tableline\tableline
 Date  & HJD             &   Rotational & Stokes        & $t_{\rm exp}$ $\times$ 4 & $\sigma_{\rm LSD} \times 10^{-5}$ & \multicolumn{1}{c}{$\langle B_{\rm z} \rangle$}   \\
            (UTC) & (2,400,000+) & phase & $V$ $Q$ $U$     &   (s)                 &                    & \multicolumn{1}{c}{(G)}           \\
\tableline 
              2012-09-25 & 56195.7684 & 0.691 & 1/1/1 & 200/400/400  & 6.2/2.1/2.1 & $124.8\pm5.8$   \\
             
              2012-09-26 & 56196.9182 & 0.862 & 1/1/1 & 200/400/400 & 5.6/2.1/2.1  & $40.7\pm5.2$     \\
             
              2012-09-27 & 56197.8587 & 0.002 & 1/1/1 & 200/400/400 &  5.7/2.0/2.0 & $66.9\pm5.2$    \\
      
              2012-09-28 & 56198.9807 & 0.168 &1/1/1 & 200/400/400 &  6.6/2.5/2.3 & $40.8\pm6.0$     \\
   
              2012-09-29 & 56199.9629 & 0.314 & 1/2/1 & 200/400/400 & 7.0/2.9$^*$/2.5 & $-76.4\pm6.4$  \\
 
	       2012-09-30 & 56200.8659 & 0.449  & 1/1/1 & 200/400/400 & 8.0/2.7/2.8 & $-72.5\pm7.3$          \\

	       2012-10-01 & 56201.7619 & 0.582  & 1/2/1 & 200/400/400 & 8.1/2.5$^*$/2.6 & $72.0\pm7.5$    \\

\tableline
	        2013-06-15 & 56459.0836 & 0.850 & 1/1/1 & 300/600/600  & 5.0/1.7/1.7 & $37.1\pm4.6$     \\
             
              2013-06-16 & 56460.0695 & 0.997 & 1/1/1 & 300/600/600 & 5.5/1.8/1.9  & $102.3\pm5.1$     \\
             
              2013-06-17 & 56461.0443 & 0.142 &  1/1/1 & 300/600/600 &  6.5/2.4/2.4 & $34.1\pm6.0$      \\
      
              2013-06-19 & 56463.0723 & 0.443 & 1/1/1 & 300/600/600 &  6.3/2.2/2.2 & $-7.1\pm5.7$       \\
   
              2013-06-21 & 56465.0526 & 0.738 & 1/1/1 & 300/600/600 & 5.3/1.9/1.9& $11.1\pm4.9$      \\
 
	       2013-06-22 & 56466.0477 & 0.886  & 1/1/1 & 300/600/600 & 5.8/2.0/1.8 & $62.2\pm5.4$       \\
	      
	       2013-06-23 & 56467.0825 & 0.040  &1/1/1 & 300/600/600 & 14.6/6.0/6.2 & $105.6\pm13.4$  \\

            2013-06-24 & 56468.0370 & 0.182  & 1/1/1 & 300/600/600 & 5.6/1.8/1.8 & $-15.3\pm5.2$    \\
	      
	      2013-06-26 & 56470.0867 & 0.486  & 1/1/1 & 300/600/600 & 5.6/1.9/1.9 & $-22.4\pm5.1$    \\
	      
	      2013-06-27 & 56471.0947 & 0.636  & 1/1/1 & 300/600/600 & 5.4/1.8/1.8 & $-26.3\pm5.1$    \\
	      
	      2013-06-28 & 56472.0748 & 0.782  & 1/1/1 & 300/600/600 & 5.5/1.8/1.9 & $25.8\pm5.1$    \\
	      
	      2013-07-01 & 56475.0615 & 0.226  & 1/1/1 & 300/600/600 & 5.3/1.8/1.8 & $-58.4\pm4.9$   \\
\tableline
\end{tabular}
\end{table*}

A commonly used technique to study stellar magnetic fields is Zeeman Doppler imaging \citep[ZDI,][]{Brown91}. This method utilizes the spectral line polarization signatures arising in the presence of a magnetic field due to the Zeeman effect. A two-dimensional vector distribution of the surface magnetic field is reconstructed by considering observations from many different rotational phases. These observations have to be obtained within a time period shorter than the evolutionary time scale of the detected magnetic field structures -- typically within several rotations.

Most cool star ZDI studies have used only circular polarization data {\citep[e.g.][]{petit2004,Marsden2006,Carroll2012}}. This is because Zeeman linear polarization signatures in spectral lines are up to 10 times weaker than circular polarization, making them more difficult to detect. In addition, interpretation of full Stokes vector spectroscopic observations requires a detailed polarized radiative transfer modeling \citep{Kochukhov2010}, which is not routinely incorporated in ZDI codes.

Most previous ZDI studies have inferred distributions of brightness spots from Stokes $I$ spectra and magnetic field from Stokes $V$ spectra using separate, inconsistent inversions.
There are some known limitations of ZDI when circular polarization is modeled without accounting for cool spots. Numerical tests by \citet{Rosen12} showed that cool temperature inhomogeneities coinciding with magnetic fields, which is common on the Sun, need to be taken into account in magnetic mapping, otherwise the magnetic field strength can be severely underestimated. A cool spot will result in an intensity decrease which will appear as a distortion in the unpolarized Stokes $I$ spectrum. The lowered intensity will also affect the polarized spectrum by causing a decrease in the amplitude of the polarization signature. The decreased amplitude will then be interpreted as a weaker field if the temperature variation is not taken into account. 

Other numerical tests \citep{Donati1997,Kochukhov02,Rosen12} consistently showed that a crosstalk between the radial and meridional field components can occur when only circular polarization is used in the magnetic inversion. This is because Stokes $V$ is only sensitive to the line-of-sight component of the magnetic field. The projected field vector of a radial or a meridional field onto the line-of-sight will always point either towards or away from the observer as the star rotates, hence their polarization signatures will behave in the same way. This is not the case for an azimuthal field vector and it is therefore easier to distinguish from the other two components. Linear polarization is, on the other hand, sensitive to the transverse component of the magnetic field vector and can hence be used to separate a radial field component from a meridional one and vice versa thus removing the crosstalk. 

A meridional field vector will be almost perpendicular to the line-of-sight, depending on the inclination of the star and the latitudinal position at the stellar surface. The projection onto the line-of-sight will hence generally be small, and the resulting Stokes $V$ signature will also be small. The meridional field component therefore has the most difficulties associated with it when it comes to reconstruction using only Stokes $V$. The projected vector onto a plane perpendicular to the line-of-sight will, on the other hand, be large, meaning the Stokes $QU$ signatures of a meridional component will be large. In general, linear polarization is more sensitive to the field orientation compared to Stokes $V$. Including linear polarization in the reconstruction process will then not only remove the crosstalk between the field components but also strengthen the meridional component and increase the overall reliability of the reconstructed magnetic field map. 

The current situation for cool star magnetic modeling is similar to the situation for Ap and Bp stars about 15 years ago. These strongly magnetic intermediate mass stars were studied using Stokes $IV$ only and were believed to have stable dipole-like fields. Then \citet{Wade2000} obtained phase-resolved, high-resolution spectra of magnetic Ap and Bp stars in all four Stokes parameters. Since then, several ZDI studies were able to recover the magnetic field topology of Ap stars {\citep[e.g.][]{Kochukhov04,Kochukhov10,Silvester2014,Rusomarov2015}} using full Stokes vector observations. These studies found that, by including linear polarization in ZDI, small-scale field structures were revealed.
 
Because of obvious advantages of the four Stokes parameter ZDI modeling, we carried out a survey trying to detect linear polarization in spectral lines of four RS~CVn stars \citep{Rosen13}. We obtained secure detections of linear polarization in all four stars. One of these targets, II~Peg, showed particularly strong linear polarization signatures and we therefore performed follow up observations of this star. Here we describe additional spectropolarimetric observations of II~Peg and perform temperature and magnetic field inversions using Stokes $IQUV$ data. This is the first four Stokes parameter ZDI analysis for a cool active star, giving us an opportunity to directly compare results of the traditional restricted Stokes $IV$ inversions with the outcome of ZDI in all four Stokes parameters. This comparison is essential for critical assessment of the reliability of previous Stokes $V$ ZDI studies of cool stars.

\section{Observations}
\label{obs}

In our previous paper \citep{Rosen13} we presented observations of II~Peg at three separate epochs. At one of these epochs II~Peg was observed for seven consecutive nights (25 September--1 October 2012) hereafter called the 2012.75 set. These data are sufficient for a ZDI inversion. We have also acquired a new set of observations during 15 June--1 July 2013 covering 12 rotational phases, hereafter called the 2013.05 set. The UT and mean Heliocentric Julian dates of all observations used in the present paper can be found in columns 1 and 2 of Table~\ref{tab1}. 

\begin{figure}
\centering
\includegraphics[scale=0.32,angle=90]{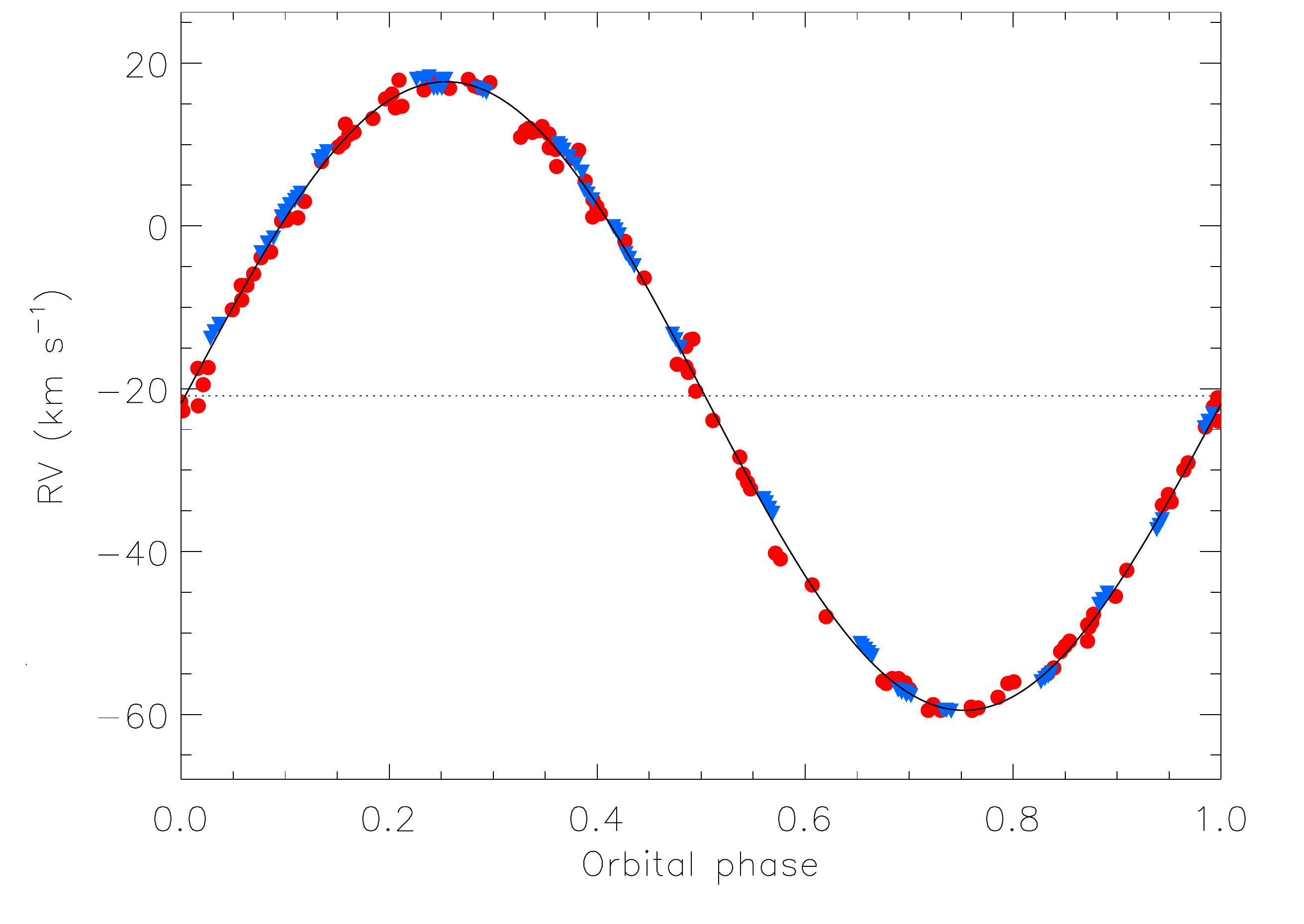}
\caption{Radial velocity of II~Peg as a function of orbital phase. The least-square fit is shown with the black solid line. The red circles are observations used by \citet{Berdyugina98} and the blue triangles are our new measurements.}
\label{orbit_fit}
\end{figure} 

\begin{table}
\caption{Revised orbital parameters for II~Peg.}
\label{orb_par}
\centering
\begin{tabular}{lll}
\tableline\tableline
Parameter & Value & Error \\
\tableline
$P$, days                &      6.7242078           & $\pm$ 0.0000068     \\
$T_{\rm conj}$, HJD      &      2448940.747         & $\pm$ 0.977   \\
$T_{\rm maxRV}$, HJD     &      2448942.428         & $\pm$ 0.977   \\
$K_{\rm 1}$, km s$^{-1}$ &      38.601              & $\pm$ 0.116  \\
$\gamma$, km s$^{-1}$    &       $-20.899$            & $\pm$ 0.083  \\
\tableline
\end{tabular}
\end{table}  

All observations have been performed at the Canada-France-Hawaii Telescope (CFHT) using the fiber-fed echelle spectrograph ESPaDOnS \citep{Donati03} in its polarimetric mode. ESPaDOnS has a wavelength coverage of 3700--10500~\AA\ and a resolving power of about $R = 65,000$. In order to derive one Stokes parameter two orthogonal polarization states are required, for example clockwise and counter clockwise circularly polarized light to obtain Stokes $V$. Both polarization states are recorded in a single exposure since the incoming beam is split into the two orthogonal states by a Wollaston prism. The beams are then transported to the spectrograph via two separate fibers where they are dispersed by the spectrograph and recorded by a 2K$\times$4.5K E2V CCD detector.

The instrument's two half-wave Fresnel rhombs can be rotated with respect to the fixed quarter-wave rhomb and the beam splitter. This feature is installed to make it possible to exchange the two beams with orthogonal polarization states between the fibers and hence also their position on the detector. By obtaining two sub-exposures with different orientations of the retarder plates, internal errors can be corrected for when a Stokes parameter is derived by combining the ratios of four spectra from two sub-exposures, as discussed by \citet{Donati97} and \citet{Bagnulo09}. In practice, each Stokes parameter observation consists of four sub-exposures. This redundancy allows the calculation of a diagnostic null spectrum \citep{Donati97}. The number of observations for each Stokes parameter and the exposure times can be found in columns 4 and 5 of Table~\ref{tab1}. The polarized spectra were reduced automatically at CFHT by the Upena pipeline using the Libre-ESpRIT software \citep{Donati97} and normalized to the continuum by performing a global smooth function fit to Stokes $I$ with the help of dedicated IDL routines. The typical peak signal-to-noise (S/N) ratio of the one-dimensional extracted spectra is about 950 per 1.8~km\,s$^{-1}$ velocity bin for the two observation epochs.

II~Peg is a spectroscopic binary displaying a single spectrum. The rotation of the primary component and the orbital motion are synchronized. The orbital radial velocity variation has to be removed prior to a ZDI analysis. To accomplish this, we measured the center-of-gravity of the mean Stokes $I$ profiles derived from all our observations of II~Peg, including the data from July 2012, December 2012 and January 2013 \citep{Rosen13}. These measurements were combined with the observations used by \citet{Berdyugina98} for their final orbital solution. A new set of orbital parameters was derived with the help of a least-squares fit assuming a circular orbit. All radial velocity measurements and the model fit can be seen in Fig.~\ref{orbit_fit} and the derived orbital parameters can be found in Table~\ref{orb_par}. The orbital period was determined to be 6.7242078~d. The rotational phase of each observation was calculated using the orbital ephemeris of $T_{\rm maxRV}=2448942.428 + 6.7242078 E$ corresponding to the time of maximum radial velocity. These phases are listed in column 3 of Table~\ref{tab1}.

\section{Multi-line analysis of polarized spectra}
\label{aps}

If a star has an average magnetic field strength of several kG, it is possible to detect polarization signatures in individual spectral lines \citep[e.g.][]{Silvester2012}. II~Peg does show circular polarization in some of the most magnetically sensitive lines. Linear polarization is weaker than circular polarization and cannot be readily seen in individual spectral lines at the S/N ratio of our observations. In order to increase the S/N ratio we applied the least-squares deconvolution (LSD) technique \citep{Donati97} using a code described by \citet{Kochukhov2010}. The goal of this multi-line method is to derive a single mean profile by combining all suitable lines in the spectrum, scaling them with line-specific weights. The weight for Stokes $I$ only depends on the depth of the line, while the weights for Stokes $VQU$ depend on depth, wavelength and effective Land\'{e} factor of the line. Since LSD assumes that all lines have similar profile shapes, lines which are significantly broader than the average (e.g. hydrogen Balmer lines, Na D) have to be excluded. 

For the purpose of LSD analysis we retrieved the line data from the Vienna Atomic Line Database \citep[{\sc vald},][]{Piskunov1995, Kupka1999}. For this extraction we adopted $T_{\rm eff}$=4750~K, $\log g=3.5$ and $[M/H]=-0.25$ according to \citet{Ottmann1998,Kochukhov2013}. From this line list we excluded all lines which have a predicted central depth less than $20~\%$ of the continuum, considering only the intrinsic broadening. We also removed lines with central wavelengths outside the 4000--8900~\AA\ region since the spectrum is noisy below 4000~\AA\ and the spectral line features are quite sparse above 8900~\AA. The LSD profiles were then calculated using 4216 lines for a total velocity range of 300 \kms\ . The resulting uncertainty per adopted 2 \kms\ velocity bin can be found in column 6 of Table~\ref{tab1}. Thanks to LSD, we achieved a mean S/N of 27000 for the Stokes $QUV$ parameters.

To find out whether a polarization signature was securely detected, we calculated the false alarm probability \citep[FAP,][]{Donati1992} for each LSD Stokes $V$, $Q$ or $U$ profile using a total velocity range of $95~$\kms\ (after correcting for the orbital motion) in order to enclose the entire profile and regions of continuum on each side. The common limit for a definite detection is FAP $<10^{-5}$ and all our observations fulfill this criterion. We also checked the FAP for all null spectra, and none of them had a FAP $<10^{-3}$ which is the limit for a marginal detection. This confirms that the detected polarization signatures are produced by the Zeeman effect in II~Peg and are not contaminated by instrumental artifacts.

Some quantitative magnetic diagnostics can be computed without invoking detailed modeling of LSD. For example, one can calculate the mean longitudinal magnetic field $\langle B_{\rm z} \rangle$ from the first moment of Stokes $V$ \citep{Kochukhov2010}. The resulting $\langle B_{\rm z} \rangle$ shows variation from $-76$ to 125~G. Compared to previous $\langle B_{\rm z} \rangle$ measurements \citep{Kochukhov2013}, this range is intermediate between a stronger longitudinal field observed in 2004--2007 and a weaker field in 2008--2010. Individual longitudinal field measurements and associated error bars are listed in column 7 of Table~\ref{tab1}.

The mean longitudinal magnetic field characterizes the sign and magnitude of the projection of the field onto the line-of-sight, integrated over the stellar disk. A complex Stokes $V$ profile indicates that the surface magnetic field structure comprises regions with different polarities at different longitudes of the star. In that case $\langle B_{\rm z} \rangle$ will be much smaller than the local field strength. Some observations, for example those with a symmetric Stokes $V$ profile, may result in $\langle B_{\rm z} \rangle$ consistent with zero, even though magnetic field is clearly present. Therefore, the diagnostic value of $\langle B_{\rm z} \rangle$ is limited for cool stars with complex fields. 

\begin{figure}
\centering
\includegraphics[scale=0.334]{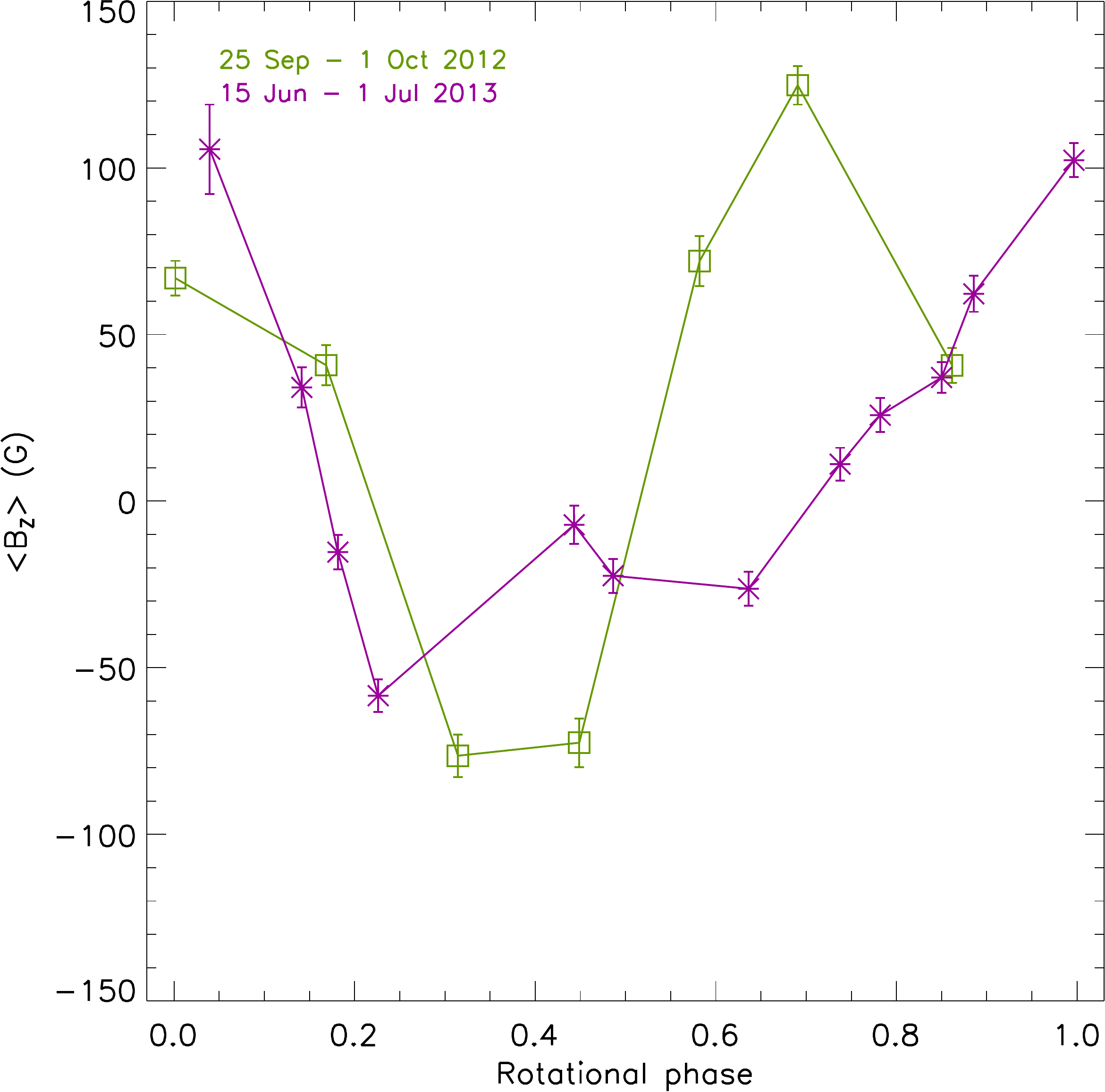}
\caption{Mean longitudinal magnetic field of II~Peg as a function of rotational phase. The green squares represent values from the 2012.75 set and the purple stars represent values from the 2013.05 set. For each point the associated error bar is shown. }
\label{bz_val}
\end{figure}

The $\langle B_{\rm z} \rangle$ of II~Peg are plotted as a function of rotational phase in Fig.~\ref{bz_val}. Some of the observations from the two epochs were taken at approximately the same phase, but the $\langle B_{\rm z} \rangle$ values in general do not agree. This suggests that the field has evolved between the two observing runs. Even though the field topology is complex, the $\langle B_{\rm z} \rangle$ variations illustrated in Fig.~\ref{bz_val} do seem fairly coherent.

\section{Zeeman Doppler imaging with LSD profiles}
\label{zdi_code}

LSD profiles are used for reconstruction of the magnetic field topology through ZDI. As discussed below, there are different approaches to ZDI, but some steps are common. The stellar surface is first divided into a grid of spatial elements. In this study we use 1176 such surface zones. Each of them are assigned an initial temperature and magnetic field strength and orientation. Synthetic local intensity and polarization line profiles are calculated for each surface zone and each rotational phase. These local Stokes profiles are then integrated over the disk and compared to the observations. The temperature and magnetic field values for each zone are updated and new local line profiles are calculated. This procedure is repeated until the deviation between the synthetic profiles and the observed profiles is below a certain threshold and the solution is no longer significantly improving.

ZDI is essentially a least-squares minimization problem where a theoretical model spectrum is fit to observations. Since ZDI is also an ill-posed problem, regularization is essential. A penalty function is therefore added in order to find the simplest possible unique solution. The contribution of the regularization function is controlled by a regularization parameter, usually denoted $\Lambda$. The penalty from the regularization should not be too small so that noise is fit or no unique solution is found, but at the same time it should not be too large so that small profile details are not fit. We determined $\Lambda$ by requiring that the contribution from the regularization function is a few times smaller than the weighted deviation between the model and the observations, and, at the same time, a good fit is obtained. The latter is achieved when the rms value of the fit is similar to the noise level of the observed profiles. Since we are reconstructing both temperature and magnetic field, we used two different regularization parameters, $\Lambda_{t}$ and $\Lambda_{f}$. We used the same regularization parameters for both the Stokes $IV$ and Stokes $IQUV$ inversions of the same observational set so that they could be easily compared. On the other hand, we used a factor 1.5--3 lower regularization for the 2013.05 set compared to the 2012.75 set. The phase coverage is better for the 2013.05 set, meaning that the solution is already better constrained and less regularization is required. 

Here, temperature is regularized with Tikhonov regularization \citep[e.g.][]{Piskunov1990} in order to suppress large gradients between neighboring surface elements. The magnetic field is represented with the help of a spherical harmonic expansion \citep{Kochukhov14}. The three vector components of the field are specified in terms of the harmonic expansion coefficients $\alpha_{l,m}$, $\beta_{l,m}$, $\gamma_{l,m}$, where $l$ is the angular degree and $m$ is the azimuthal order of each mode. The coefficients $\alpha_{l,m}$ represent the radial poloidal component, $\beta_{l,m}$ the horizontal poloidal component and $\gamma_{l,m}$ the horizontal toroidal component. The magnetic inversion is regularized with a harmonic penalty function $\sum_{l,m}l^2(\alpha_{l,m}^2+\beta_{l,m}^2+\gamma_{l,m}^2)$, which suppresses unnecessary high-order terms. To find an upper limit for $l$, we kept increasing $l_{\rm{max}}$ until the highest $l$-mode would hold an insignificant fraction of the total energy. This resulted in an upper limit of $l_{\rm{max}}=20$ for this study.       

In general, it is preferable to use individual spectral lines for reconstruction of the stellar temperature distribution and magnetic field geometry. Several lines with well known line-parameters and with different magnetic and temperature sensitivity can be chosen from the observed spectrum and modeled through detailed polarized radiative transfer calculations to constrain the solution \citep[e.g.][]{Rosen12}. The situation is quite different when dealing with four Stokes parameter LSD profiles. There is no possibility to constrain the solution by using lines with different behaviors. Furthermore, each average Stokes profile is a mean of thousands of spectral line signatures and cannot be easily and uniquely assigned a set of parameters that would reproduce its response to temperature inhomogeneities and magnetic field \citep{Kochukhov2010}.

\subsection{Single-line approximation of LSD profiles}

The traditional approach of using LSD profiles for ZDI is to treat them as single lines with a set of average parameters. One method is to fit Stokes $I$ with a Gaussian function by adjusting its width and strength. Then the weak-field approximation is employed to calculate Stokes $V$ as the derivative of Stokes $I$ \citep[e.g.][]{Marsden2011}. 

Another common approach is to use a Milne-Eddington model atmosphere in order to solve the polarized radiative transfer equation analytically \citep[e.g.][]{Brown91}. Here there are also some free parameters, in particular the linear source function slope, characteristics of the absorption and anomalous dispersion profiles, and the Zeeman splitting pattern, that need to be set.

Yet another approach is to approximate the local LSD profiles by solving the polarized radiative transfer equation using realistic model atmospheres. In this case a theoretical model profile is typically calculated using the parameters derived by averaging over all lines of the most common ion in the LSD line mask. The Stokes $I$ and $V$ observations of II~Peg have previously been modeled using this method  \citep{Kochukhov2013}.

All these methods rely on a number of more or less restrictive approximations and require adjusting several line parameters. There are no straightforward, well-documented procedures of how to perform this adjustment. Again, an LSD profile is an average over thousands of spectral lines and, in general, its behavior cannot be accurately reproduced with a single spectral line. Numerical tests by \citet{Kochukhov2010} showed that the single-line approximation of an LSD profile only holds for Stokes $IV$ if the field strength is below $\sim$2kG. However, the single-line approach could not approximate the LSD Stokes $QU$ profiles at all. Therefore, magnetic inversions in all four Stokes parameters cannot rely on any of the previously used single-line LSD profile approximations.

\subsection{Grids of pre-calculated local LSD profiles}

In this study of II~Peg we have applied a new method, implemented in a new ZDI code called {\sc inversLSD}, developed by \citet{Kochukhov14} in order to overcome the problems inherent in single-line approximations of LSD profiles. In this method we avoid assigning any specific line parameters to LSD profiles when comparing observations and theoretical models. Instead, a grid of local synthetic LSD profiles is calculated from the full polarized spectrum synthesis using the same line mask as was used for analysis of the observations. Each such synthetic LSD profile corresponds to a specific temperature, magnetic field strength, limb angle and magnetic field vector orientation with respect to the line of sight. 

Here we have used 19 {\sc marcs} model atmospheres \citep{Gustafsson2008} with temperatures from 3000--6000~K in order to take temperature inhomogeneities into account. A metallicity of [M/H]=\,$-0.25$, surface gravity of $\log g$=\,3.5, microturbulent velocity of $\xi_{\rm t}=2.0$~km\,s$^{-1}$, and radial-tangential macroturbulent broadening of $\zeta_{\rm t}=4.0$~km\,s$^{-1}$ were adopted to match the stellar parameters of II~Peg \citep{Ottmann1998,Kochukhov2013}. In order to cover a sufficiently wide range of magnetic field configurations, we have calculated Stokes $IQUV$ profiles corresponding to magnetic field strengths between 0 and 3000~G with a step of 100~G, 15 limb angles $\delta$, spaced equidistantly in $\cos \delta$ and 15 magnetic field vector inclinations $\theta$, with respect to the line-of-sight, also spaced equidistantly in $\cos \theta$, resulting in 132525 unique LSD profiles. 

The first step of every iteration was to obtain local LSD profiles for each surface element. This was done by a linear interpolation in the pre-calculated grid of LSD profiles, significantly accelerating the inversion procedure. The Stokes $QU$ profiles are transformed on-the-fly according to the orientation of the local field vector in the plane of the sky. The next step was to integrate all the local LSD Stokes profiles over the stellar disk. The disk-integrated synthetic LSD profiles were then compared to the observed LSD profiles directly, without making any assumptions about the behavior of the LSD profiles. This is one of the advantages of this approach compared to the traditional single-line approximation of LSD profiles. Another crucial advantage of this method is that it can be applied to all four Stokes parameters regardless of the magnetic field strength.   

The {\sc inversLSD} code used here largely builds on the {\sc invers13} code \citep{Kochukhov2012,Kochukhov2013}. The main difference is that it interpolates within precomputed local profiles instead of performing polarized radiative transfer calculations on-the-fly.

In addition to local Stokes profiles, other parameters necessary for ZDI include the projected rotational velocity $v_{\rm e}\sin i$, the inclination angle $i$, and the azimuthal angle $\theta$ of the rotational axis (projected on the plane of the sky). The latter parameter is required when dealing with linear polarization since it defines orientation of the magnetic field vector in the plane perpendicular to the line-of-sight. We used $i=60^\circ$ \citep{Berdyugina98,Frasca2008,Hackman2012,Kochukhov2013} and determined $v_{\rm e}\sin i=23$~\kms\ and $\theta=75^\circ$ by minimizing the $\chi^2$ of the fit to the observed Stokes $IQUV$ LSD profiles.

\subsection{Combining individual-line Doppler imaging with LSD ZDI}

As explained above, it is preferable to employ individual spectral lines for reconstruction of the surface temperature and magnetic field distributions. Stokes $VQU$ should be modeled simultaneously and therefore we do not have any other choice but to use their LSD profiles since no linear polarization signatures are visible in individual lines. Stokes $I$, on the other hand, does show clear distortions in individual spectral lines due to temperature inhomogeneities. Thus, we took advantage of the information in individual spectral lines aiming to obtain temperature distributions at the same level of accuracy as in previous temperature DI studies of II~Peg \citep[e.g.][]{Hackman2012}. We used three Fe~{\sc i} lines at 5987.1, 6003.0 and 6024.1~\AA\ for reconstruction of temperature maps. These lines have different magnetic sensitivities and two of them are slightly blended with several other lines while one is unblended. We performed all temperature inversions using the {\sc invers13} code, taking into account the magnetic field during temperature reconstructions.

Since we used two separate codes, the temperature and magnetic field inversions were carried out separately. However, in each magnetic inversion a fixed inhomogeneous temperature distribution was taken into account and vice versa. The first step was to perform a temperature inversion using the three Fe~{\sc i} lines assuming a null magnetic field and $T_{\rm{eff}}$\,=\,4750~K as an initial guess for the temperature. At the next step we obtained a magnetic map from the LSD Stokes $VQU$ profiles using the previously derived temperature distribution as a fixed temperature structure. The resulting magnetic map was then used as a fixed magnetic field distribution for another temperature inversion where the initial temperature distribution was now that which was derived in the first temperature inversion. This new temperature distribution was inserted as a fixed temperature for another magnetic field derivation using the previously derived magnetic field structure as an initial guess. Then another temperature inversion was performed and then, as a final step, another magnetic inversion. Thus, we iterated both temperature and magnetic maps three times. There was no reason to continue iterations further because there were no significant changes in the temperature distribution between the second and third temperature inversion, hence the magnetic field topology also remained almost unchanged between the two last magnetic inversions. 

Since we have observations in all four Stokes parameters, we have the capability to compare Stokes $IV$ inversions to Stokes $IQUV$ inversions. We therefore produced two sets of maps for each data set; one where we only used Stokes $IV$ for the reconstruction and one where we used all four Stokes parameters.

\begin{figure*}
\centering
\includegraphics[scale=0.5]{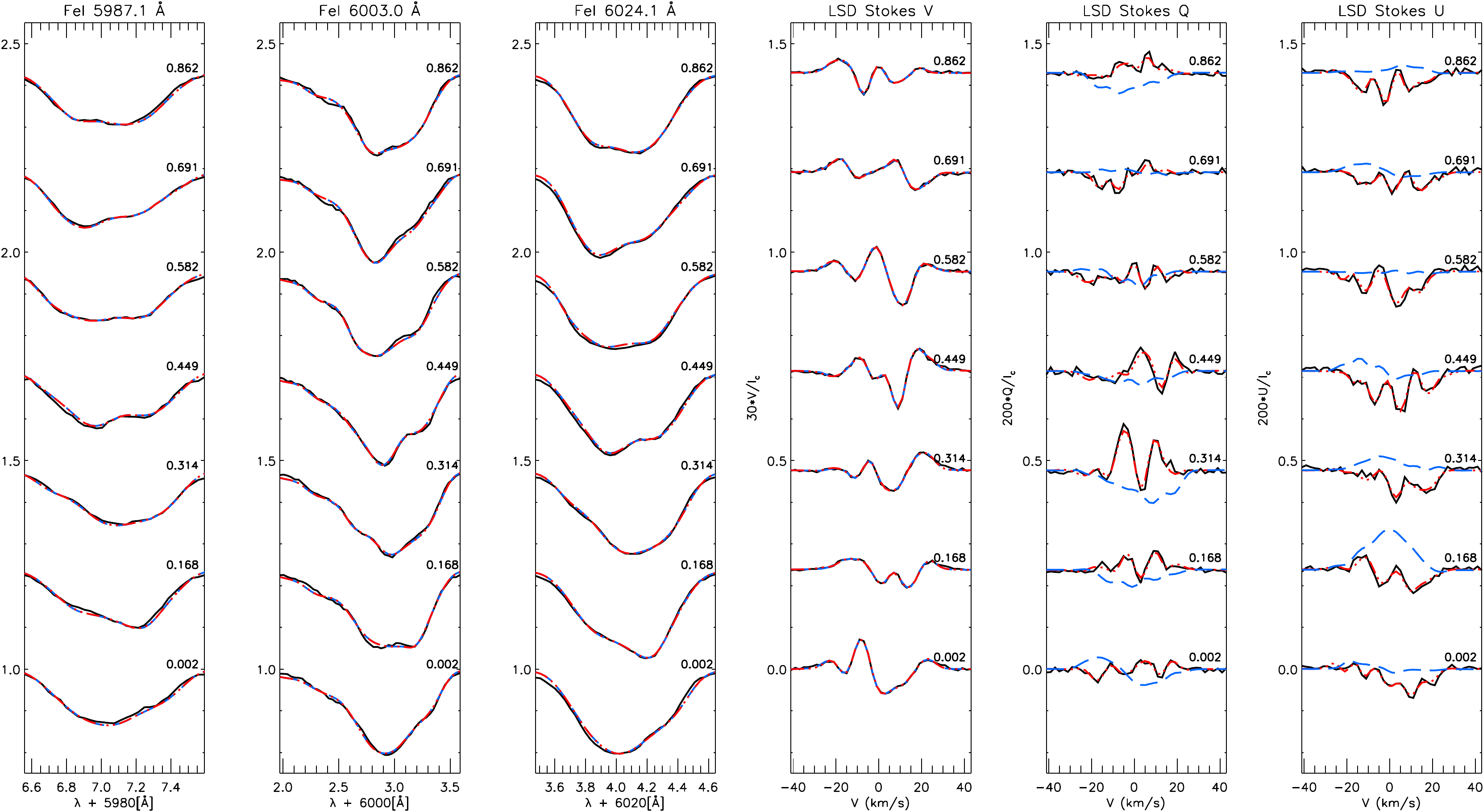}
\caption{The observed and model Stokes $IQUV$ profiles of II~Peg for the 2012.75 data set. The first three panels show the Stokes $I$ profiles of the Fe~{\sc i} lines used in the inversions. The next three panels show the Stokes $VQU$ LSD profiles. All spectra are offset vertically. The Stokes $Q$ and $U$ profiles are magnified by a factor of 200 and the Stokes $V$ profiles are magnified by a factor of 30 relative to Stokes $I$. The black solid lines represent the observations. The blue dashed line corresponds to the model profiles for the Stokes $IV$ inversion. The red dash-dotted lines represent the model profiles for the full Stokes vector inversion.}
\label{lines12}
\end{figure*}

\begin{figure*}
\centering
\includegraphics[scale=0.63,angle=90]{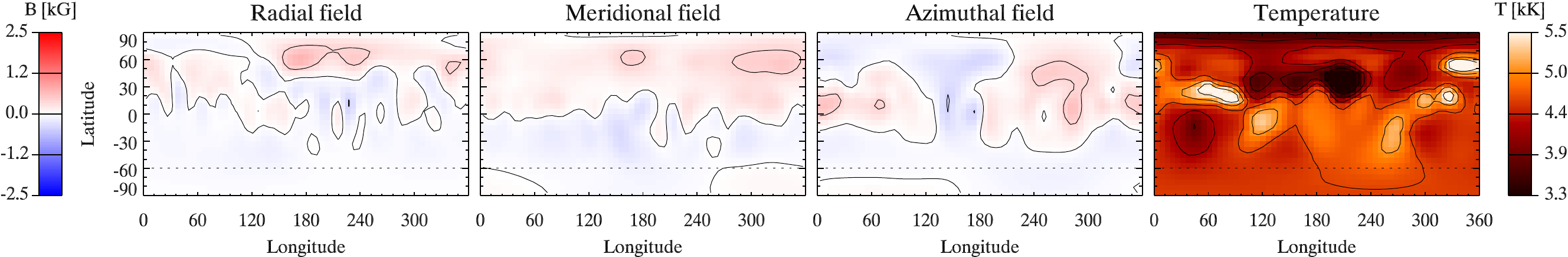} \\
\includegraphics[scale=0.63,angle=90]{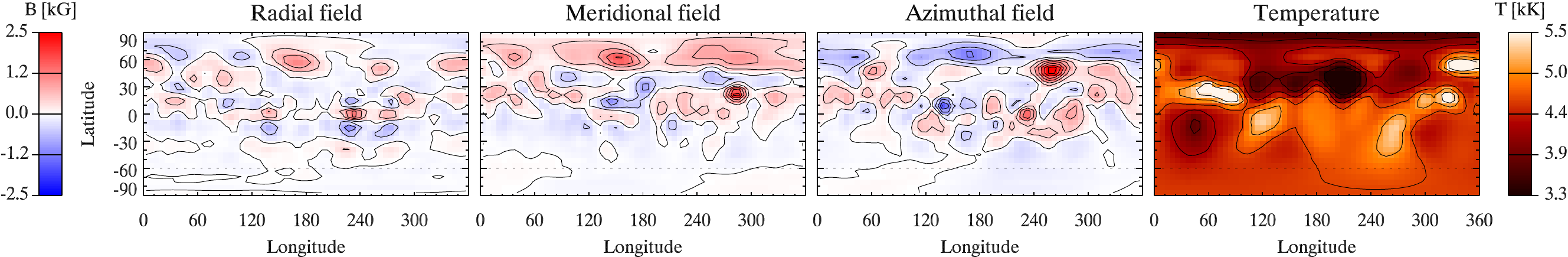} \\
\includegraphics[scale=0.63,angle=90]{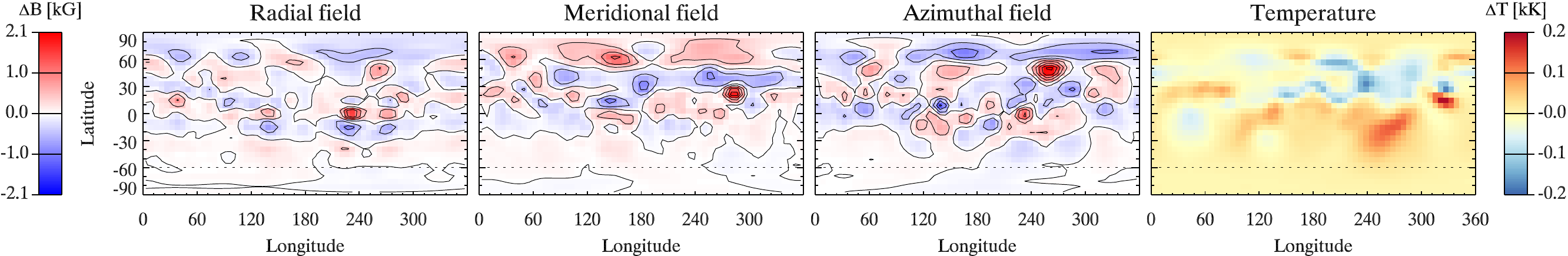}
\caption{Rectangular projections of the magnetic and temperature maps recovered for the 2012.75 data set. The top row corresponds to the Stokes $IV$ inversion. The middle row presents results of the Stokes $IQUV$ reconstruction. The bottom row shows the difference between the two inversions. The contour lines are plotted with a step of 400~G in magnetic and 400~K in temperature maps.}
\label{rec12}
\end{figure*}

\begin{figure*}
\centering
\includegraphics[width=0.58\textwidth,angle=270]{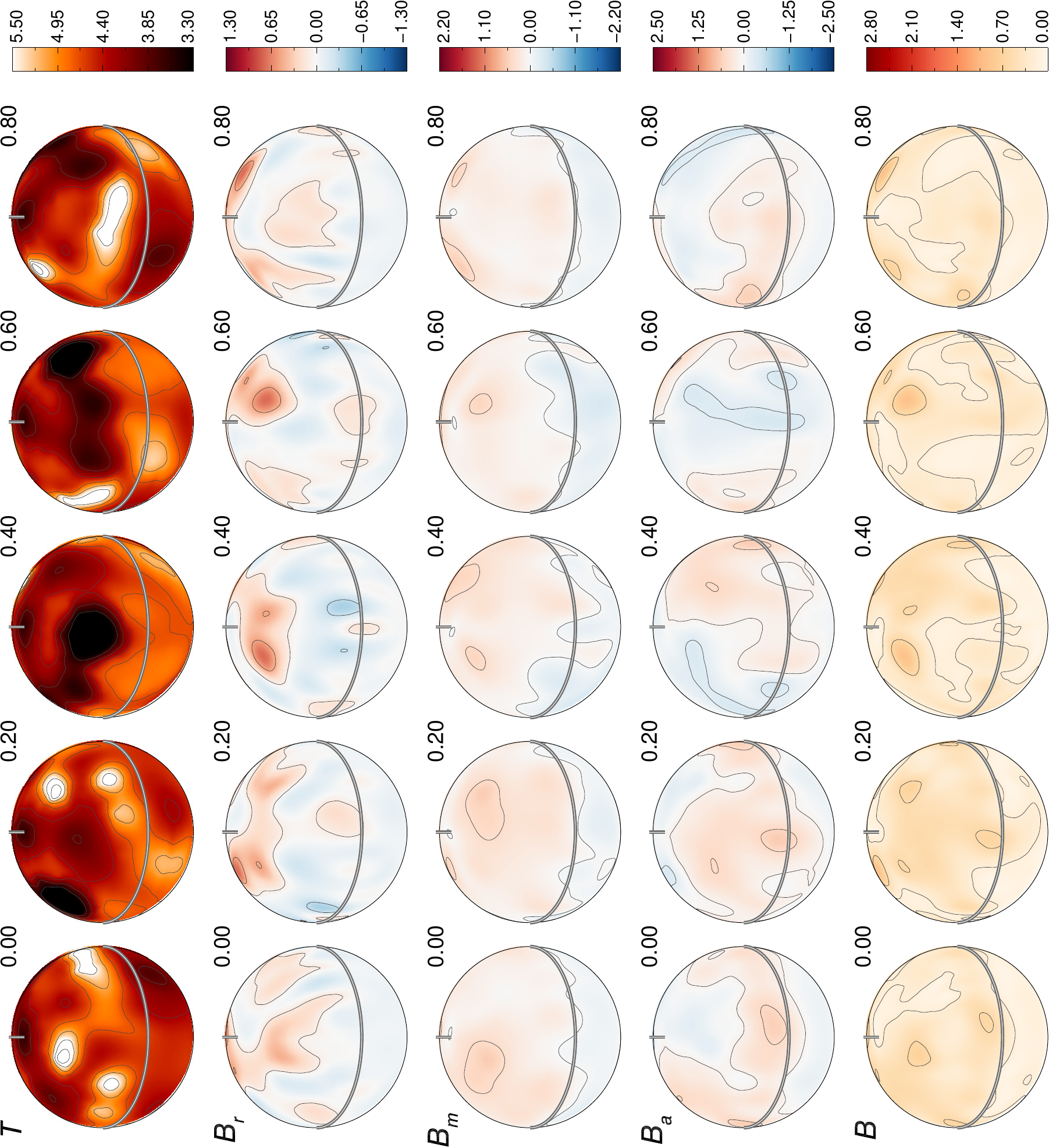}  
\caption{Spherical projections of the magnetic and temperature maps recovered for the 2012.75 data set using Stokes $IV$ parameters. The rows of spherical plots show, from top to bottom, distributions of temperature, radial, meridional, azimuthal field components, and the field modulus. The star is shown at five rotational phases indicated above each panel. The double line indicates positions of the rotational equator. The vertical bar indicates location of the rotational pole.}
\label{sph12_2}
\end{figure*}

\begin{figure*}
\centering
\includegraphics[width=0.58\textwidth,angle=270]{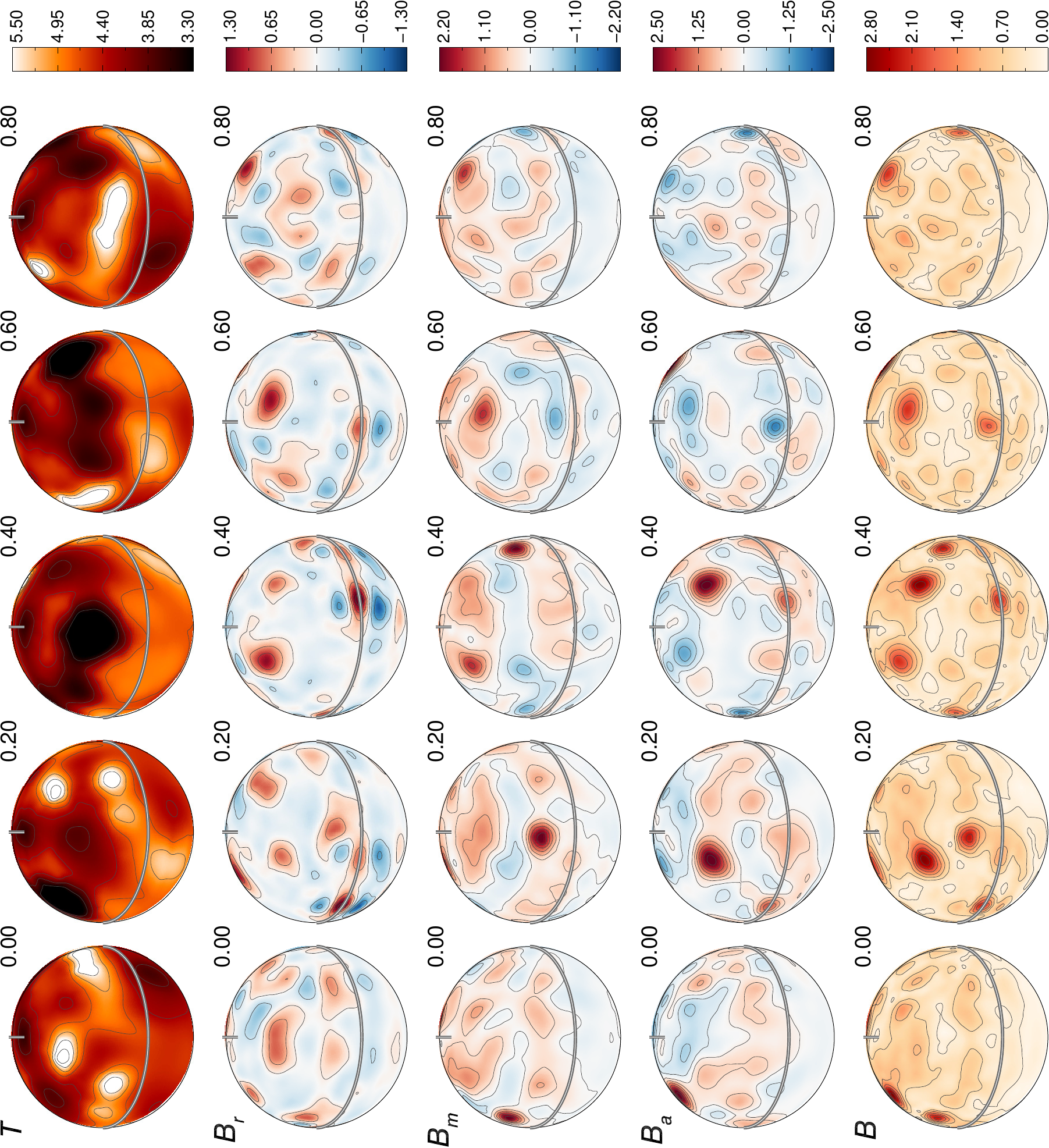}
\caption{Same as Fig.~\ref{sph12_2} but for the four Stokes parameter inversion.}
\label{sph12_4}
\end{figure*}

\begin{figure*}
\centering
\includegraphics[scale=0.41]{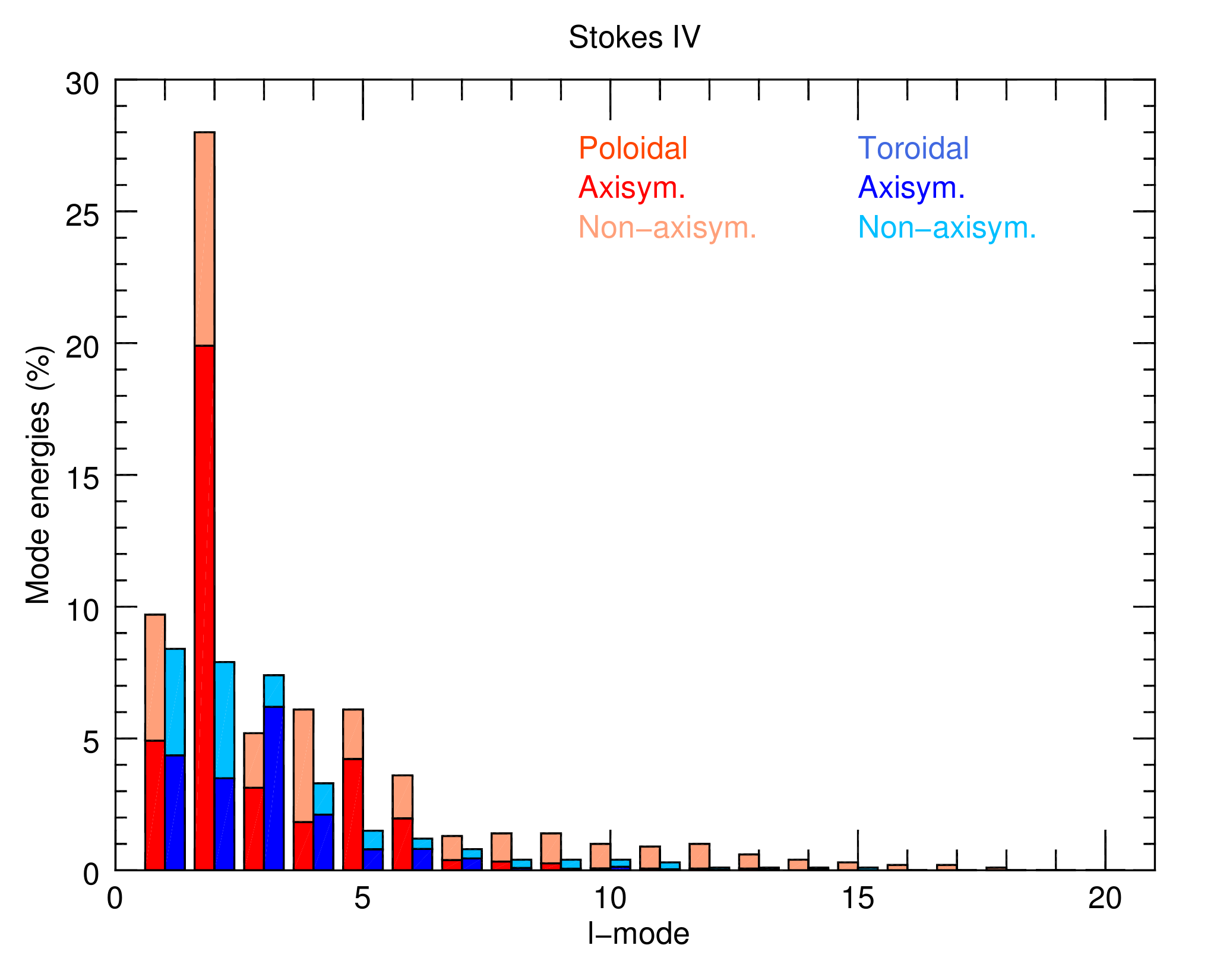} 
\includegraphics[scale=0.41]{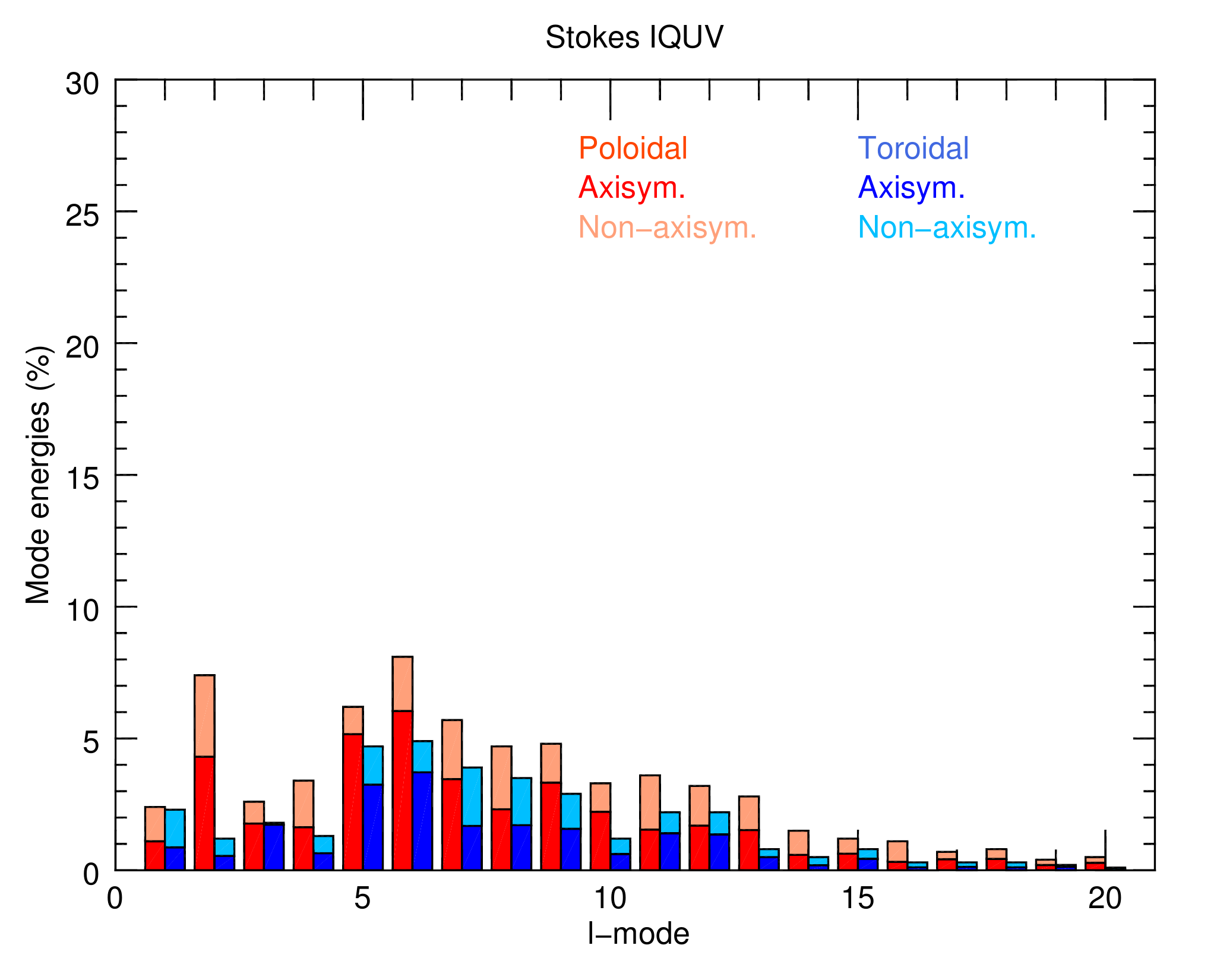}
\caption{Distribution of the magnetic field energy between different spherical harmonic modes for the 2012.75 data set. The left panel shows results of the Stokes $IV$ inversion. The right panel corresponds to the Stokes $IQUV$ inversion. In each panel the red bars represent the poloidal component and the blue bars correspond to the toroidal component. Each bar is also divided into an axisymmetric (darker shades) and non-axisymmetric (lighter shades) part. 
}
\label{sep12_mode}
\end{figure*}

\section{Results}
\label{res}

\subsection{2012.75 data set}

The observed polarized LSD line profiles are complex and change in shape and amplitude from one rotational phase to another, as can be seen in Fig.~\ref{lines12}. The Stokes $I$ profiles of the three Fe~{\sc i} lines show clear variability due to the presence of temperature inhomogeneities. Also displayed in Fig.~\ref{lines12} are the model profiles for the inversions based on the Stokes $IV$ data and on all four Stokes parameters. Rectangular projections of the reconstructed distributions of the three field components and temperature are presented in Fig.~\ref{rec12}. The spherical projections of the same maps as well as the field modulus ($B=\sqrt{B_{\rm r}^2+B_{\rm m}^2+B_{\rm a}^2}$) distribution are shown in Figs.~\ref{sph12_2} and \ref{sph12_4}. To get a more detailed view of the discrepancies between the two inversions, we subtracted the values of the Stokes $IV$ maps from the Stokes $IQUV$ maps. The resulting difference maps are displayed at the bottom of Fig.~\ref{rec12}.

\begin{table}
\small
\caption{Root-mean-square values of the magnetic field components and the field modulus for different ZDI inversions.}
\label{tab2}
\centering
\begin{tabular}{cccccc}
\tableline\tableline
Data set & Used Stokes & $\langle B_{\rm r} \rangle$ & $\langle B_{\rm m} \rangle$ & $\langle B_{\rm a} \rangle$ & $\langle B \rangle$ \\
               & parameters     &    (G)                                   &    (G)                                    &    (G)                                   &    (G) \\
\tableline
2012.75      &      $IV$                    &  130                                   &        169                              &        197                               &   290   \\
              &      $IQUV$                 &  238                                    &        327                               &        361                               &   543   \\
\tableline
2013.05  &      $IV$                    &  197                                   &         242                               &        307                               &   438   \\
              &      $IQUV$             &  256                                    &        440                              &        384                              &   638   \\              
\tableline
\end{tabular}
\end{table}  

One of the first things to notice in Fig.~\ref{lines12} is a mismatch between the Stokes $QU$ profiles corresponding to the reconstructed magnetic field in the Stokes $IV$ inversion (blue dashed lines) and the observed Stokes $QU$ profiles (black solid lines). They differ in sign, strength and complexity. However, when linear polarization is taken into account in the magnetic field reconstruction process, the observed $QU$ profiles are reproduced quite accurately by the model profiles (red dash-dotted lines), with a rms of about $3.4\cdot10^{-5}$. Using the values from column 6 of Table~\ref{tab1} the mean $\sigma_{\rm LSD}$ for Stokes $QU$ can be calculated to be about $2.4\cdot10^{-5}$.

The discrepancy between the two inversions visible in the Stokes $QU$ line profiles corresponds directly to dramatic differences in the respective magnetic field distributions (see Figs.~\ref{rec12}--\ref{sph12_4}). When linear polarization is taken into account in the ZDI inversion, strong features, up to 1.3, 2.1 and 2.5~kG for the radial, meridional and azimuthal field components respectively, are recovered. This should be compared to a maximum of just about 0.7, 0.6 and 0.6~kG for the same three components when only circular polarization is considered. The increase in strength can also be seen in the root-mean-square (rms) values of the radial, meridional and azimuthal components, and the field modulus listed in Table~\ref{tab2}. All components are significantly increased, by 83--93\%, when linear polarization is included. The radial component seems to be the weakest of the three components and the azimuthal is the strongest. At the same time, the largest increase is found for the meridional component. 

We also calculated the total magnetic energy by integrating the field modulus over the stellar surface. It follows the same pattern of a significant increase when all four Stokes parameters are modeled by ZDI. The magnetic energy in the case of the Stokes $IQUV$ inversion is 3.5 times higher compared to the Stokes $IV$ inversion.

Additionally, we looked at temperatures and magnetic field strengths of individual surface elements. It turns out that about 30\% of the surface elements have opposite polarity and about 26\% are stronger in the Stokes $IV$ maps compared to the Stokes $IQUV$ maps.

\begin{figure*}
\centering
\includegraphics[scale=0.8]{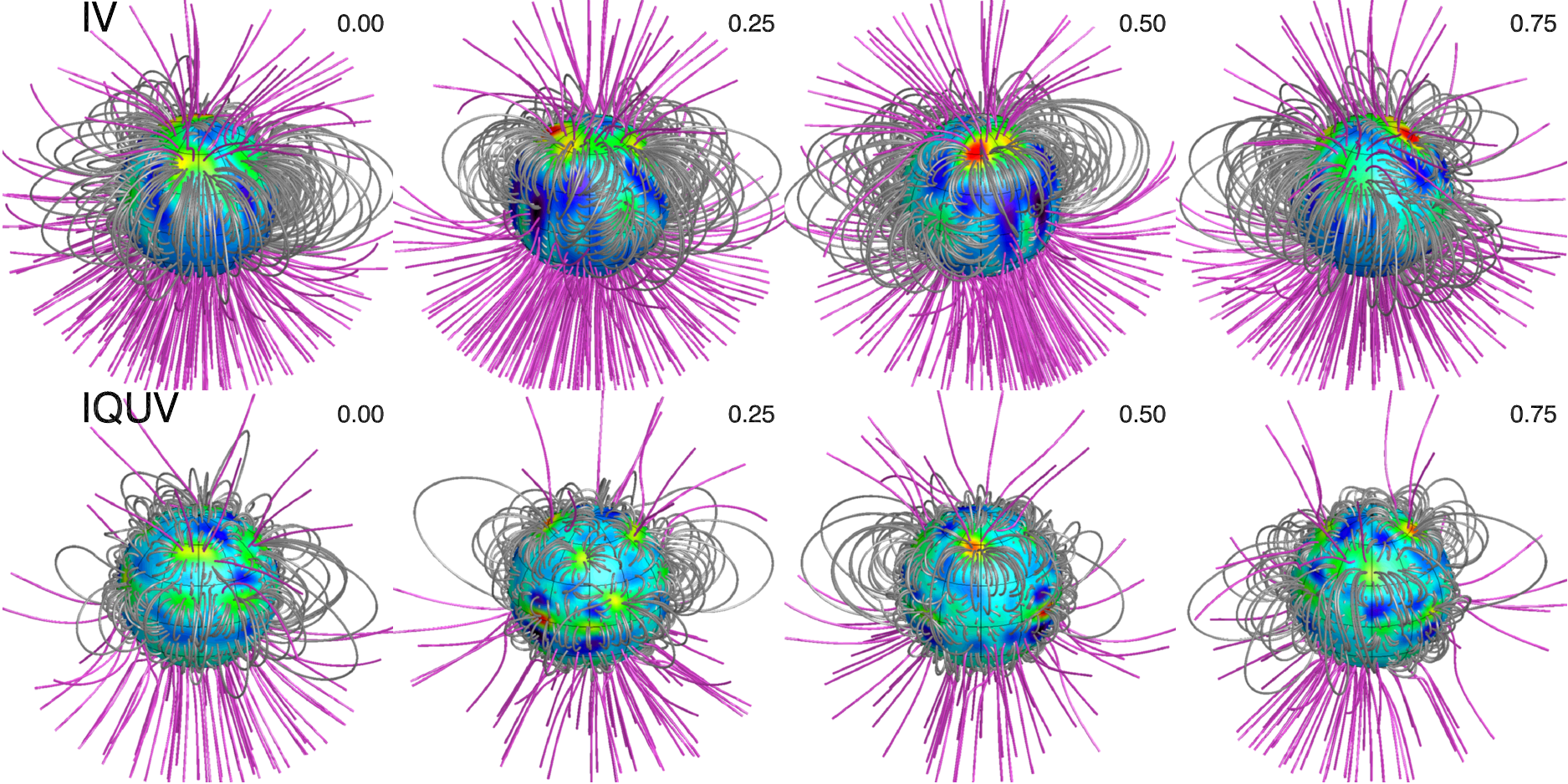} 
\caption{Three-dimensional rendering of the extended magnetic field topology of II~Peg inferred with the potential field extrapolation from the Stokes $IV$ (upper row) and Stokes $IQUV$ (lower row) ZDI maps obtained for the 2012.75 data set. The star is shown at four rotational phases indicated above each panel. The open and closed magnetic field lines are shown with different color. The underlying spherical map corresponds to the stellar surface distribution of the radial magnetic field component.}
\label{pot12_lines}
\end{figure*} 

\begin{figure*}
\centering
\includegraphics[scale=0.5,angle=270]{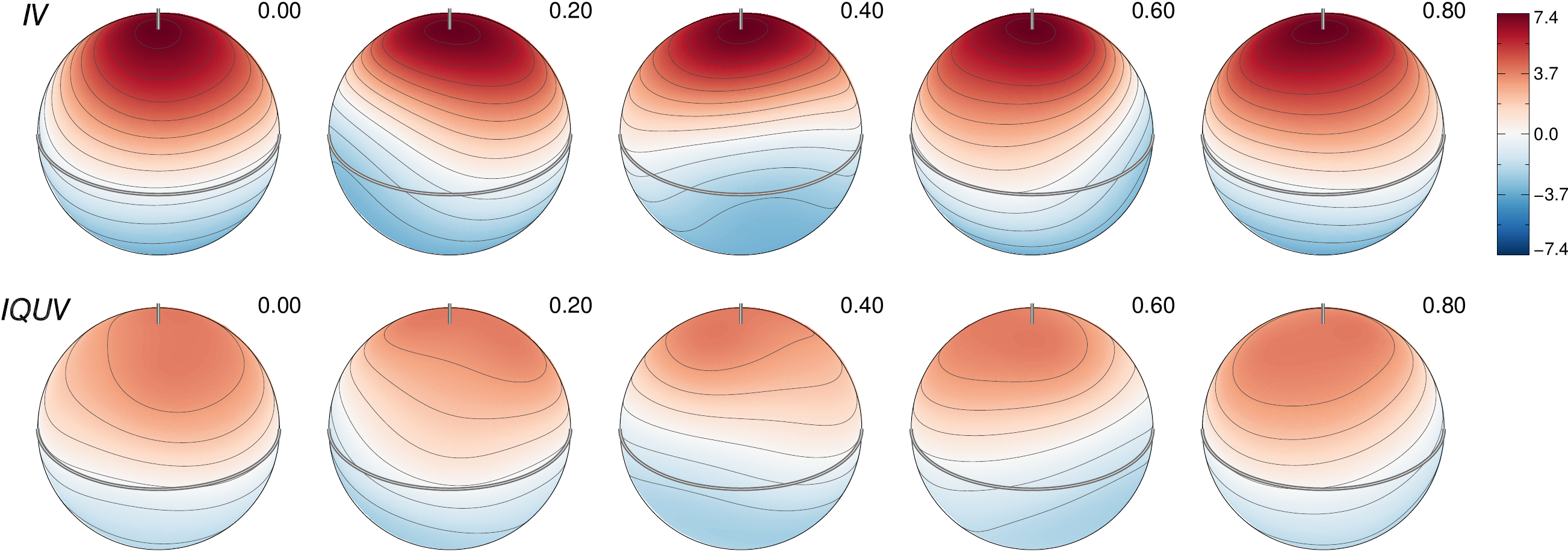}
\caption{Distribution of the radial magnetic field at the source surface (i.e the Alfv\'{e}n radius) found with the potential field extrapolation from the Stokes $IV$ (upper row) and Stokes $IQUV$ (lower row) ZDI maps obtained for the 2012.75 data set. The star is shown at five rotational phases indicated above each panel. The field strength is illustrated using the color scale (with the legend at right, in units of gauss.)}
\label{pot12_br}
\end{figure*}

The field is a {\it{significantly}} more complex in the four Stokes inversion, as can be directly seen in the ZDI maps. Some features change polarity depending on how many Stokes parameters are used. With Stokes $IV$, the visible pole has a predominantly positive radial field, but when linear polarization is taken into account it is the other way around. Almost the entire upper stellar hemisphere has a positive meridional field when only Stokes $IV$ are modeled, but that is not the case in the four Stokes parameter map. A very strong magnetic spot also appears at latitude 10--30$^\circ$ and longitude 270--300$^\circ$ in the Stokes $IQUV$ map. The azimuthal map follows the same pattern as the other two magnetic components. It also becomes more complex and detailed when all four Stokes parameters are considered. A prominent azimuthal field spot appears at latitude 30--60$^\circ$ and longitude 240--280$^\circ$. The changes in strength of the recovered magnetic field distribution are also reflected in the field modulus map shown in Figs.~\ref{sph12_2} and \ref{sph12_4}.

Even though the magnetic field topologies obtained in the two inversions are clearly different, the Stokes $V$ model profiles are indistinguishable and fit equally well the observed LSD $V$ profiles with a rms of about $8.4\cdot10^{-5}$. This vividly illustrates a non-uniqueness of the Stokes $V$ magnetic inversion and its tendency to miss small-scale magnetic fields. The rms value is also slightly larger than the mean $\sigma_{\rm LSD}$ of about $6.7\cdot10^{-5}$, calculated from column 6 in Table~\ref{tab1} for Stokes $V$.

\begin{table*}
\caption{Distribution of poloidal, toroidal, axisymmetric and non-axisymmetric field energies.}
\label{tab_mode}
\centering
\begin{tabular}{cccccc}
\tableline\tableline
Data set & Used Stokes           & $E_{\rm pol}/E_{\rm tor}$ & $E_{\rm a}/E_{\rm n}$ & $E_{\rm a}/E_{\rm n}$      & $E_{\rm a}/E_{\rm n}$ \\
               & parameters             &    (\% $E_{\rm tot}$)            &    (\% $E_{\rm tot}$)        &    (\% $E_{\rm pol}$)       &    (\% $E_{\rm tor}$) \\
\tableline
2012.75      &      $IV$                   &  67.5/32.5                             &      56.2/43.8              &     55.4/44.6                  &   57.6/42.4   \\
              &      $IQUV$                   &  64.5/35.5                           &        60.0/40.0             &     60.7/39.3                   &  58.8/41.2  \\
\tableline
2013.05  &      $IV$                    &  56.7/43.3                             &       60.8/39.2               &       55.3/44.7                 &   67.9/32.1  \\
              &      $IQUV$                &  67.4/32.6                            &      64.6/35.4                 &     65.5/34.5                   &  62.5/37.5 \\              
\tableline
\end{tabular}
\end{table*}

The increased complexity of the magnetic field topology can be quantified by considering the distribution of magnetic field energy between different harmonic modes. In Fig.~\ref{sep12_mode} the distribution of energies of the poloidal/toroidal components for each $l$-mode is displayed in red and blue respectively, and in column 3 in Table~\ref{tab_mode} the ratios over all $l$-modes combined are listed. For the Stokes $IV$ inversion, the quadrupolar ($l$=2) component is the largest, holding 35.9\% of the total mode energy. Only 4.2\% of the energy is found in modes with $l > 10$. In contrast, the energy is much more distributed across the components, and significant contributions are found at much higher $l$ values when Stokes $QU$ are included in ZDI, with 23.3\% of the total energy distributed among $l > 10$. There is a slight maximum of 13.0\% at $l=6$. In both cases the field is predominantly poloidal, but slightly less so when linear polarization is taken into account (see column 3 in Table~\ref{tab_mode}).

Analysis of the energy distribution of different harmonic modes shows that the four Stokes parameter inversion recovers more energy in the high-$l$ modes but a significantly weaker field in the $l\le5$ harmonic components than the Stokes $IV$ reconstruction. Thus, the global field geometry is not correctly retrieved by the Stokes $IV$ ZDI.

In order to obtain more information about the field structure we calculated the relative contribution of the axisymmetric and non-axisymmetric harmonic components. In the context of ZDI, all modes with $m < l/2$ are defined as axisymmetric and modes with $m \geq l/2$ are considered as non-axisymmetric \citep{Fares2009}. In both the Stokes $IV$ and Stokes $IQUV$ inversion, the field is mainly axisymmetric with a similar ratio (see column 4 in Table~\ref{tab_mode}). Similar ratios between axisymmetric and non-axisymmetric contributions are found for the poloidal and toroidal components individually (see columns 5--6 in Table~\ref{tab_mode}). The axisymmetric part for each poloidal and toroidal component is illustrated in Fig.~\ref{sep12_mode} in dark red and dark blue respectively.

Results of ZDI inversions are often used to assess an extended magnetic field topology with the goal to study the impact of magnetic field on the stellar mass loss, coronal emission, etc. \citep[e.g.][]{Hussain2002}. We calculated a three-dimensional magnetic field structure for both types of ZDI inversions with the help of the potential source surface extrapolation method \citep{Jardine02}. These calculations use the ZDI radial field component as one boundary condition. Another is set by assuming that the field lines become purely radial at the Alfv\'{e}n radius $R_{\rm s}$, where the magnetic energy is equal to the kinetic energy of the stellar wind. We assumed $R_{\rm{s}}=3R_{\star}$, which is similar to values used in previous studies of cool active stars. The results of the potential field extrapolation are illustrated in Figs.~\ref{pot12_lines} and \ref{pot12_br}. In the first figure the open and closed magnetic field lines are shown for both the Stokes $IV$ map and the Stokes $IQUV$ map. There seems to be more open field lines for the Stokes $IV$ inversion. The second figure shows the radial magnetic field map at $R_{\rm{s}}$. Here it can be seen that, although the magnetic field topologies are qualitatively similar, the field corresponding to the Stokes $IV$ inversion is much stronger compared to field extrapolated from the Stokes $IQUV$ inversion results. The total magnetic energy at $R_{\rm s}$ is about 2.5 times higher for the Stokes $IV$ case, indicating that an extrapolation from such an inversion significantly overestimates the size of the stellar magnetosphere.

The temperature map reconstructed in the Stokes $IV$ inversion is similar to the results of the four Stokes parameter reconstruction, with a maximum difference of 200~K. Some features are hotter than the stellar $T_{\rm eff}=4750$~K. The origin of these hot spots is unclear, but similar features were found in previous DI studies of II~Peg \citep{Hackman2012}. The visible pole of the star is cooler by about 1000~K relative to the stellar $T_{\rm eff}$. There is also a large spot around latitude 20--60$^\circ$ and longitude 180--240$^\circ$ with a temperature of about 3400~K. No immediate connection can be found between these spots and any prominent magnetic features. The field strength of the coolest spot is only about 0.1~kG in the Stokes $IV$ map and about 0.2~kG in the Stokes $IQUV$ map. However, the strongest azimuthal spot in the Stokes $IQUV$ map has a temperature of about 4100~K. 

\begin{figure*}
\centering
\includegraphics[scale=0.5]{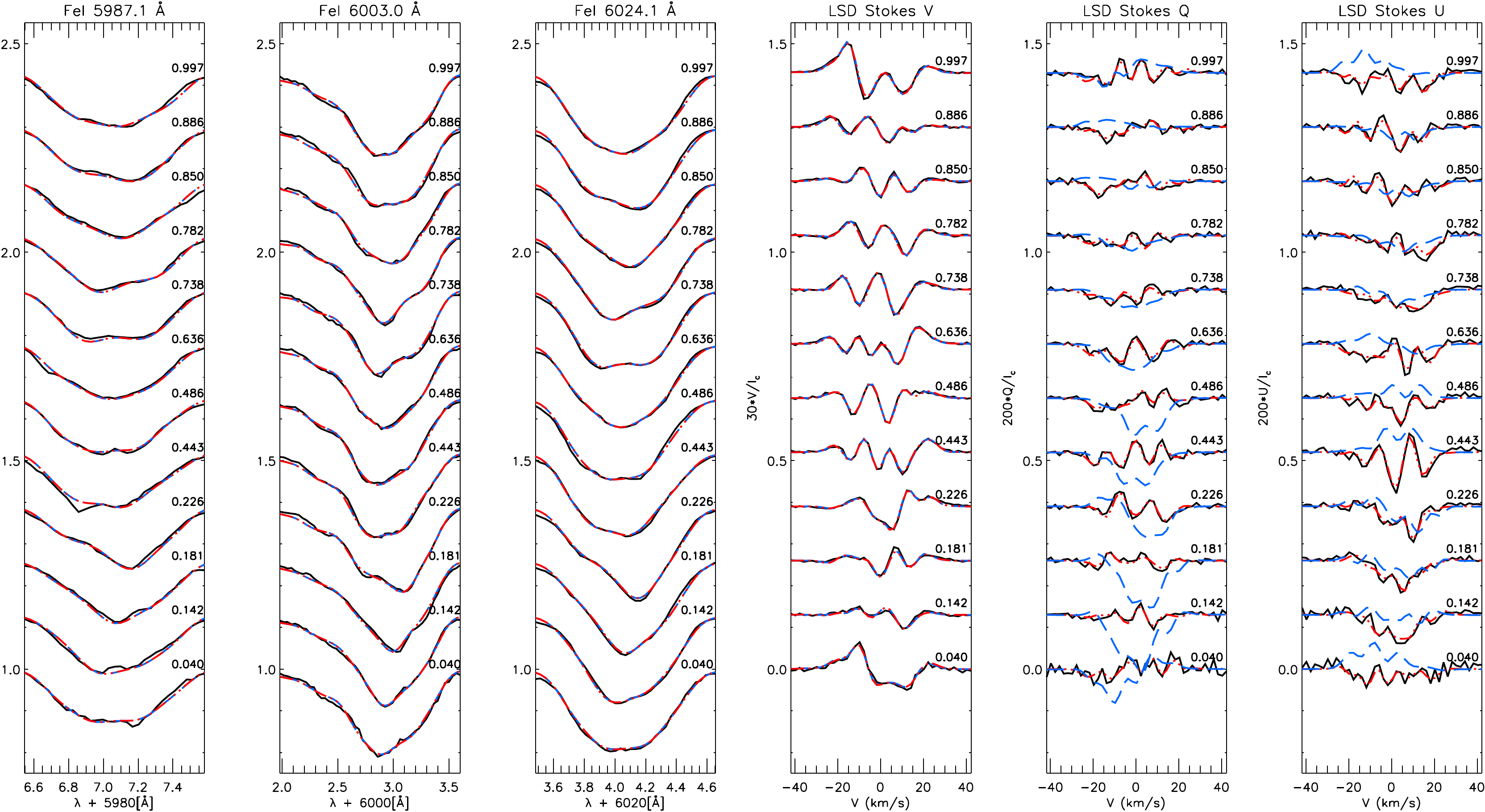}
\caption{Same as in Fig.~\ref{lines12} but for the 2013.05 data set.
}
\label{lines13}
\end{figure*}

\begin{figure*}
\centering
\includegraphics[scale=0.63,angle=90]{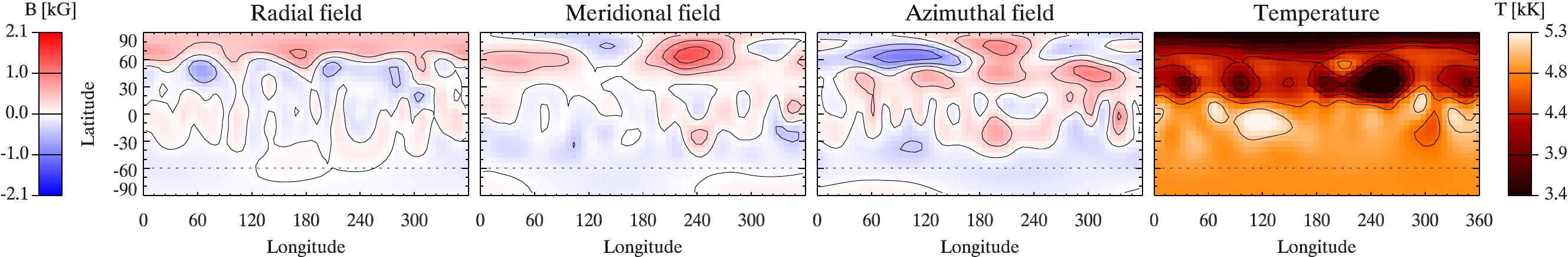} \\
\includegraphics[scale=0.63,angle=90]{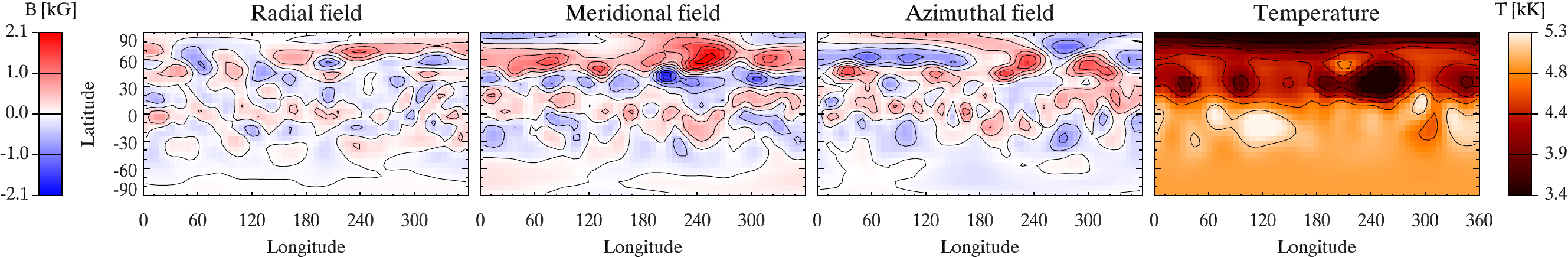} \\
\includegraphics[scale=0.63,angle=90]{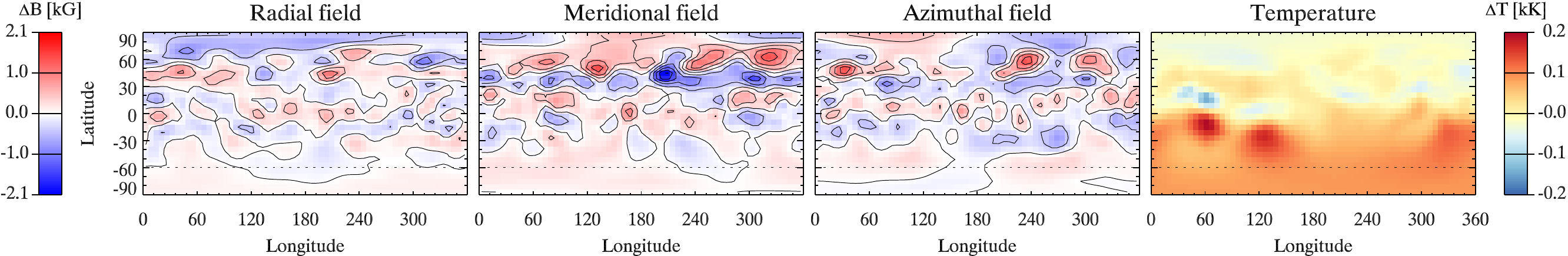}
\caption{Same as in Fig.~\ref{rec12} but for the 2013.05 data set.
}
\label{rec13}
\end{figure*}

\begin{figure*}
\centering
\includegraphics[width=0.58\textwidth,angle=270]{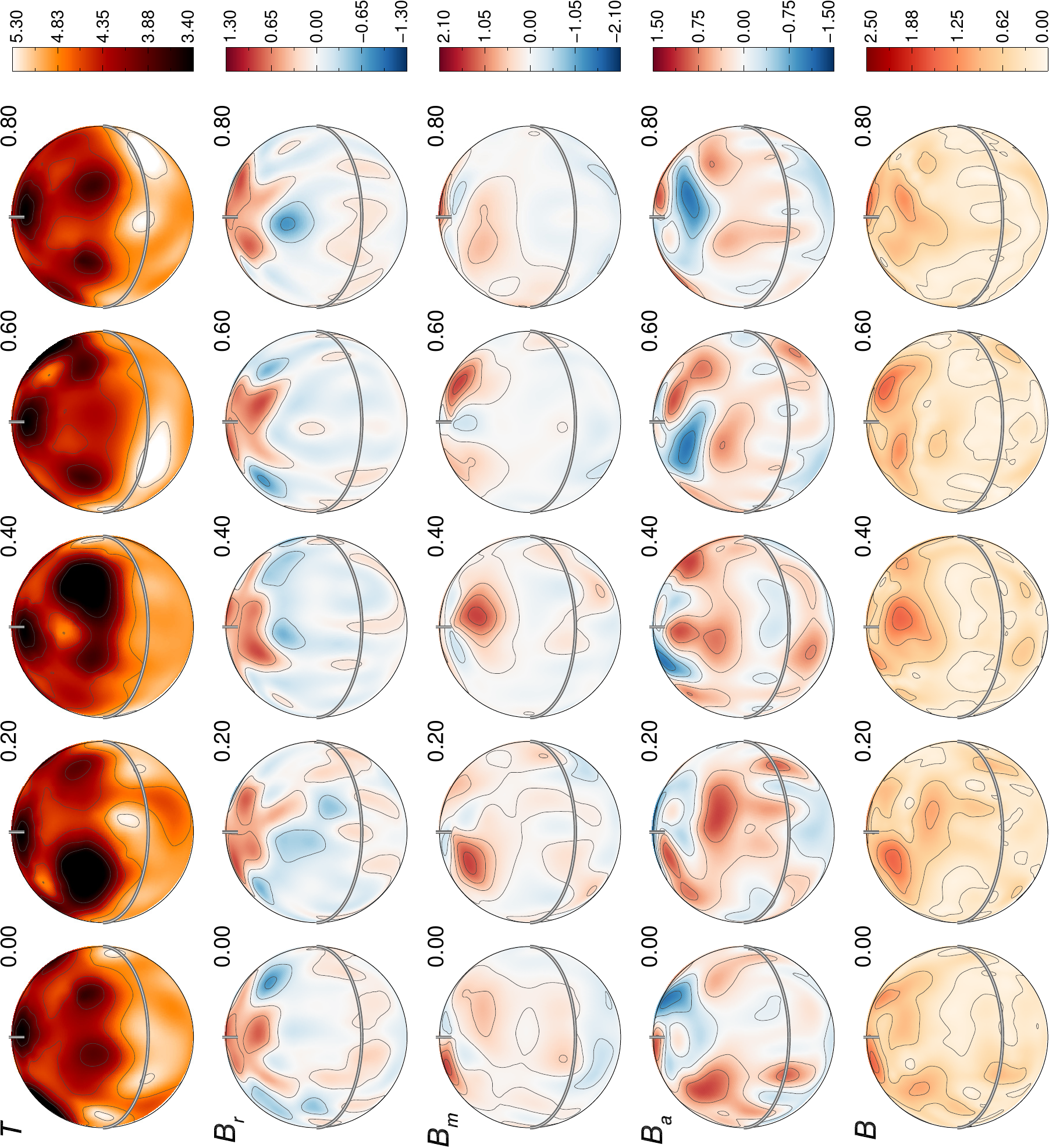} 
\caption{Same as in Fig.~\ref{sph12_2} but for the 2013.05 data set.
}
\label{sph13_2}
\end{figure*}

\begin{figure*}
\centering
\includegraphics[width=0.58\textwidth,angle=270]{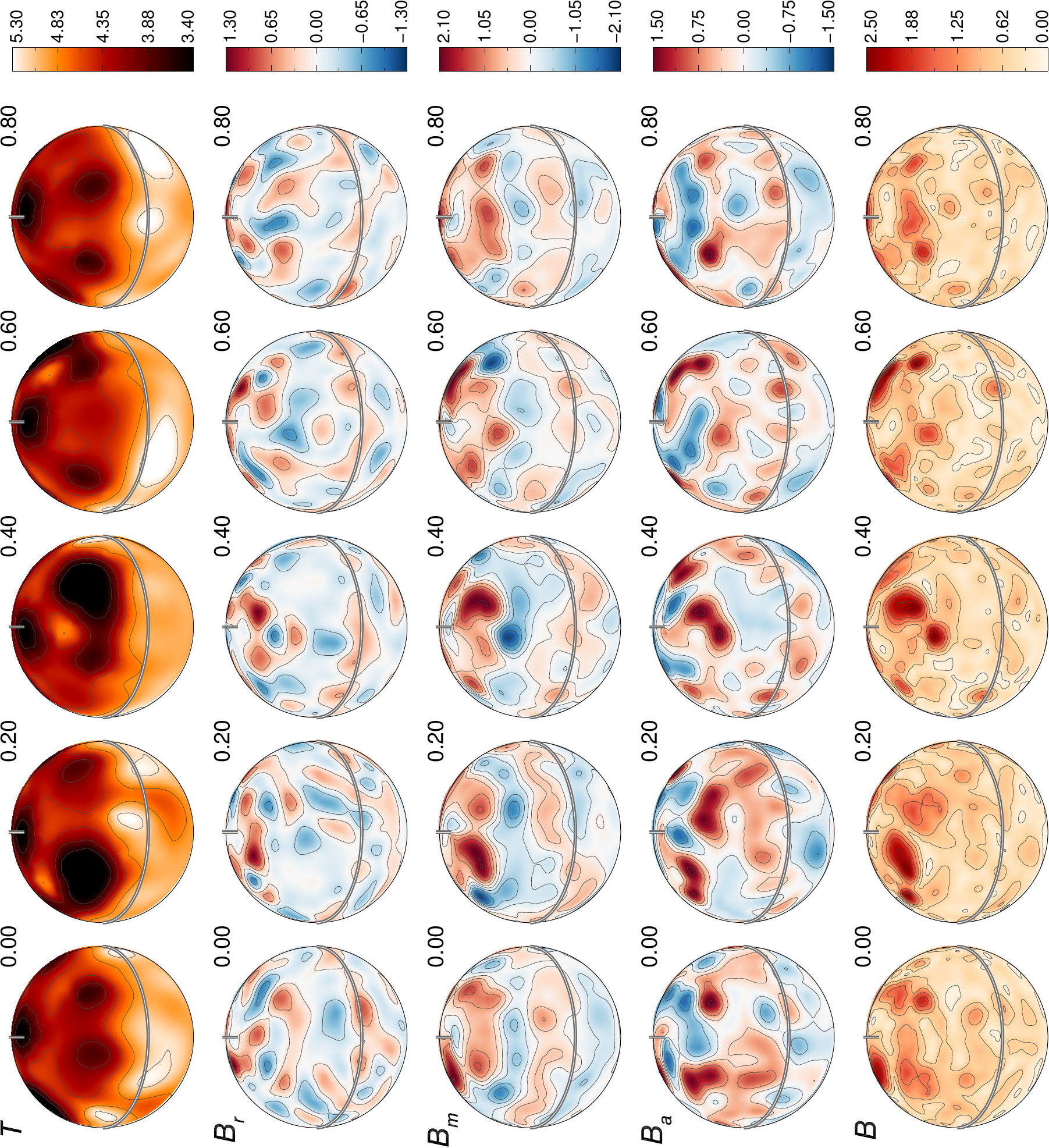}
\caption{Same as in Fig.~\ref{sph12_4} but for the 2013.05 data set. 
}
\label{sph13_4}
\end{figure*}

\begin{figure*}
\centering
\includegraphics[scale=0.41]{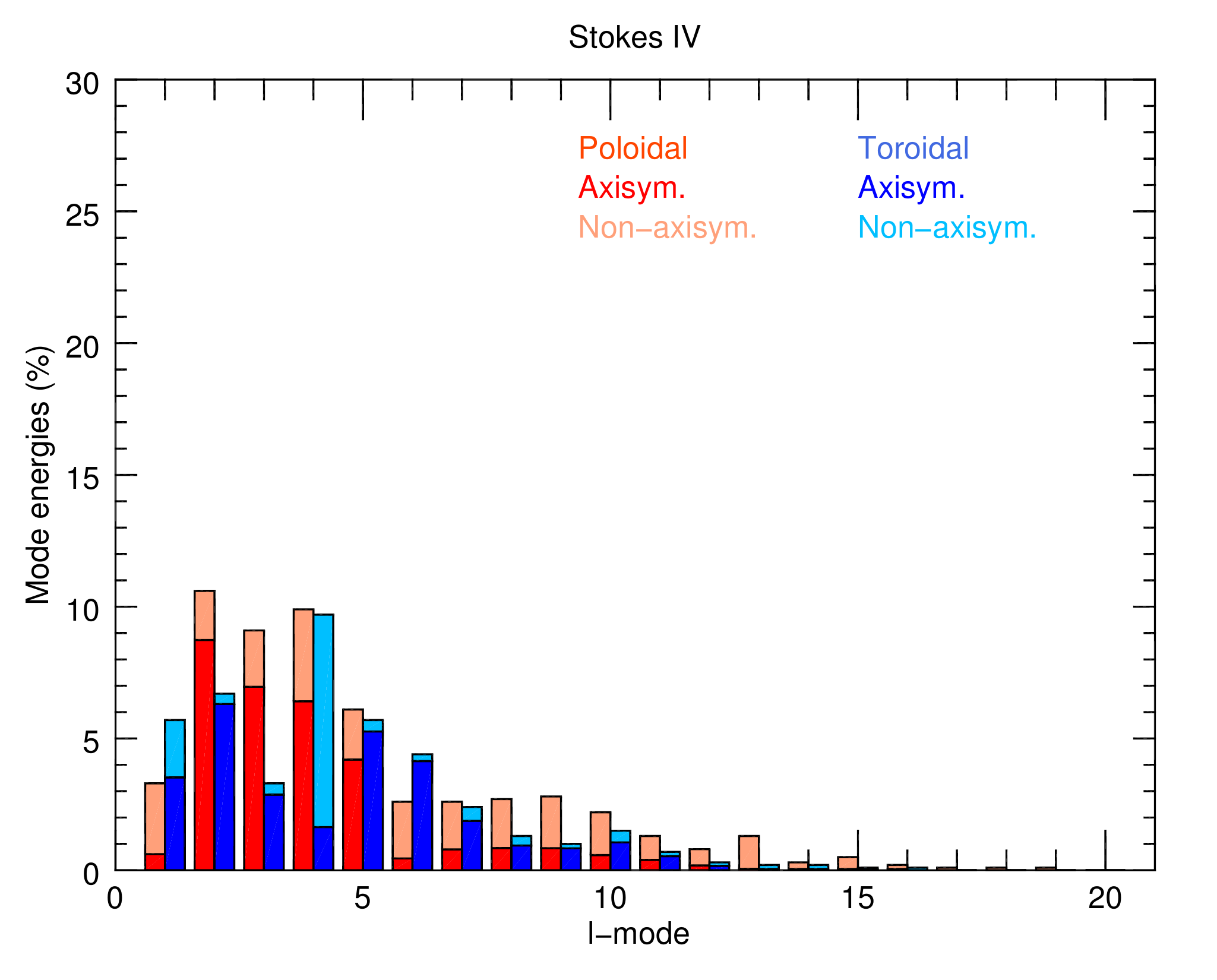} 
\includegraphics[scale=0.41]{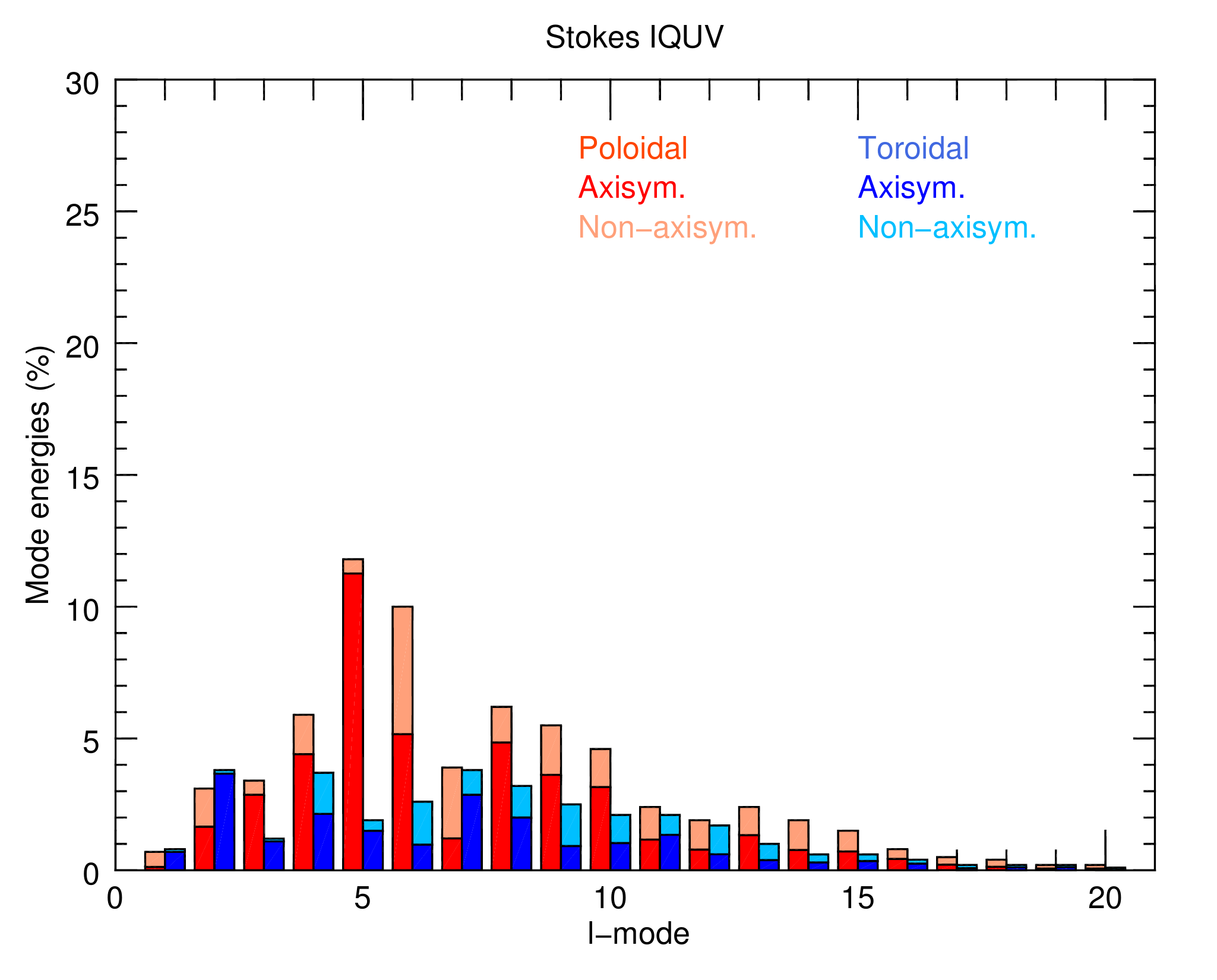}
\caption{Same as for Fig.~\ref{sep12_mode} but for the 2013.05 data set.
}
\label{sum13_mode}
\end{figure*}

\begin{figure*}
\centering
\includegraphics[scale=0.8]{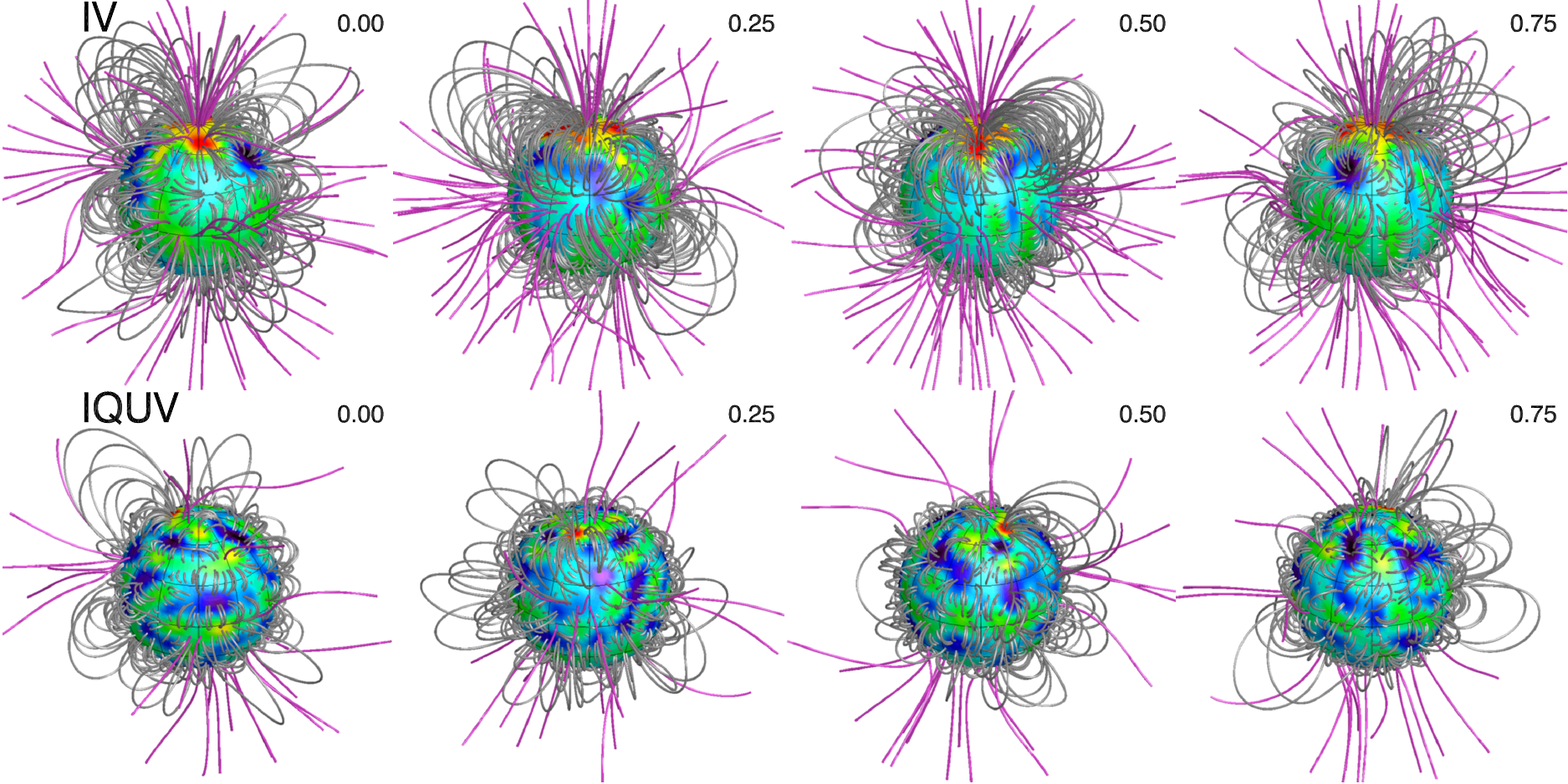}   
\caption{Same as for Fig.~\ref{pot12_lines} but for the 2013.05 data set.
}
\label{pot13_lines}
\end{figure*} 

\begin{figure*}
\centering
\includegraphics[scale=0.5,angle=270]{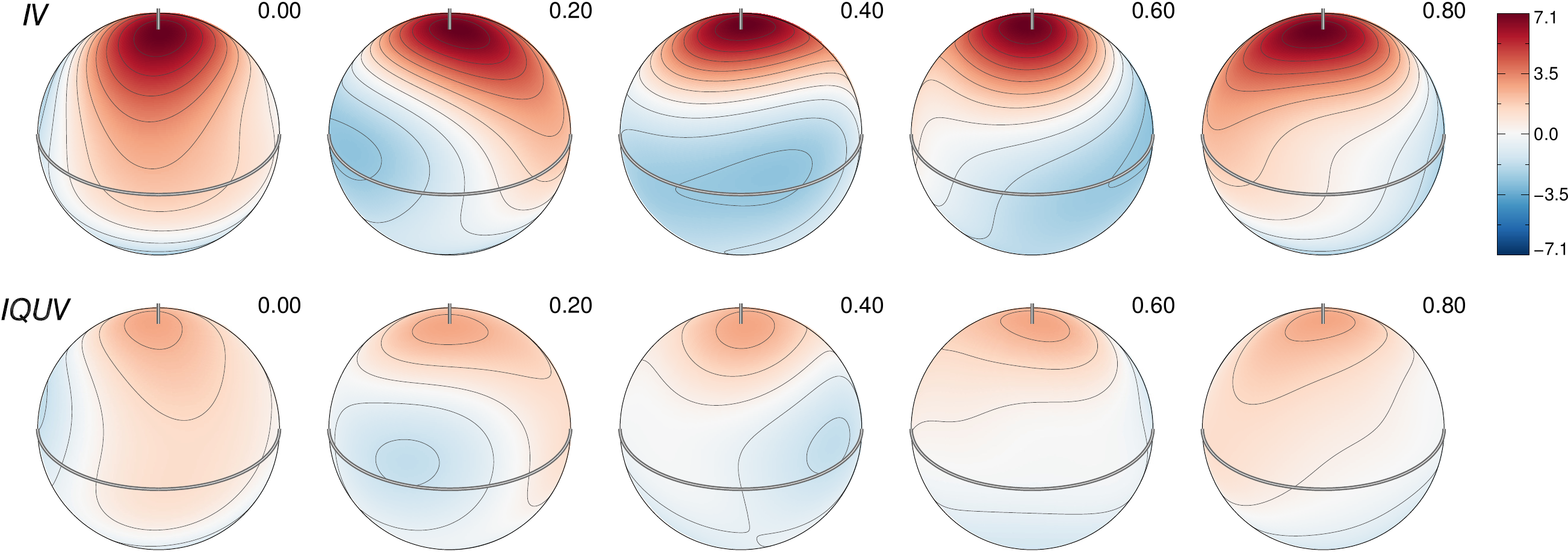}
\caption{Same as for Fig.~\ref{pot12_br} but for the 2013.05 data set.
}
\label{pot13_br}
\end{figure*} 

\subsection{2013.05 data set}

The observed LSD line profiles in the Stokes $VQU$ parameters are complex and show clear polarization signatures at all rotational phases. The three Fe~{\sc i} lines show signs of temperature inhomogeneities, as can be seen in Fig.~\ref{lines13}.  

Once again, the Stokes $QU$ model profiles corresponding to the magnetic field reconstructed in the Stokes $IV$ inversion do not fit the observed Stokes $QU$ profiles. However, the ZDI code is able to reproduce the complex observed Stokes $QU$ profiles (in addition to the $V$ profiles) with a rms of about $3.6\cdot10^{-5}$ when linear polarization data are included in the inversion. This value compares reasonably well to the mean $\sigma_{\rm LSD}$ of about $2.3\cdot10^{-5}$ for Stokes $QU$, calculated from the values listed in column 6 of Table~\ref{tab1}. 

The differences in the model Stokes $QU$ profiles for the two inversions are reflected in the magnetic field topologies, as illustrated in Figs.~\ref{rec13}--\ref{sph13_4}. The strongest field structures in both inversions are found for the meridional component. In the Stokes $IV$ inversion the maximum strength is about 0.8, 1.4 and 1.1~kG for the radial, meridional and azimuthal component respectively. In the Stokes $IQUV$ inversion the maxima are increased to about 1.3, 2.1 and 1.5~kG for these three components. This increase is also reflected in the rms values (Table~\ref{tab2}). The largest increase (82\%) is once again found for the rms field of the meridional component and the smallest (25\%) for the rms of the azimuthal component. The total magnetic energy is about 2.1 times higher in the Stokes $IQUV$ inversion compared to the map recovered in the Stokes $IV$ inversion.

To further investigate the discrepancies between the two inversions we looked at the difference between the corresponding vector magnetic maps (bottom panel in Fig.~\ref{rec13}). For about 25\% of the surface elements, magnetic field values found in the Stokes $IV$ inversion are of the same sign but stronger than the corresponding Stokes $IQUV$ values and about 34\% of the surface elements have opposite polarities in the two inversions.

As can be seen from the rectangular (Fig.~\ref{rec13}) and spherical (Figs.~\ref{sph13_2} and \ref{sph13_4}) maps, the magnetic field becomes considerably more structured in the ZDI carried out in all four Stokes parameters. The visible pole has a positive radial field when only Stokes $IV$ profiles are used to reconstruct the magnetic field but when linear polarization is included, the pole is no longer entirely positive. Instead, a few negative magnetic features are recovered in between the positive field areas. The strongest meridional field spot in the Stokes $IV$ map is divided into two even stronger spots of negative and positive polarity in the Stokes $IQUV$ map. The azimuthal component also becomes more structured, with smaller features of a higher strength emerging in the Stokes $IQUV$ inversion case.

The overall increase of the field complexity can also be seen in the distribution of the harmonic mode energies (Fig.~\ref{sum13_mode}). The total energy in the $l$-modes larger than 10 is 6.3\% for the Stokes $IV$ inversion and 20.8\% for the Stokes $IQUV$ inversion. The $l=4$ mode contains 19.6\% of the total energy and the quadrupolar component is a close second with 17.3\% for the Stokes $IV$ case. The first five $l$-modes hold 70.2\% of the total magnetic energy. On the other hand, for the Stokes $IQUV$ inversion the $l=5$--6 modes dominate with 13.7\% and 12.6\% of the total energy respectively. 

As in the analysis of the 2012.75 data set, the field is found to be predominantly poloidal in both inversions but slightly more so in the four Stokes parameter inversion (see column 3 in Table~\ref{tab_mode}). Again, both field distributions are mainly axisymmetric (see column 4 in Table~\ref{tab_mode}).

Even though the magnetic field topologies are different, the Stokes $V$ model profiles are very similar and fit the observed Stokes $V$ profiles  equally well, with a rms of about $1.0\cdot10^{-4}$. The mean $\sigma_{\rm LSD}$ of Stokes $V$ calculated from the values in column 6 in Table~\ref{tab1} is about $6.4\cdot10^{-5}$.

The extended field topology obtained with the help of potential field extrapolation from the ZDI results is illustrated in Fig.~\ref{pot13_lines} (magnetic field lines) and Fig.~\ref{pot13_br} ($B_{\rm r}$ at the source surface). There are more open field lines in the Stokes $IV$ case compared to the Stokes $IQUV$ case and, as we have seen for the inversions with the 2012.75 data set, the field at the source surface is stronger. Here the difference is larger: the total magnetic field energy at $R_{\rm{s}}$ is about 5.3 times larger in the Stokes $IV$ inversion.

The temperature maps are similar in the two inversions with a maximum difference of about 180~K (see lower panel in Fig.~\ref{rec13}). The pole, once again, has a temperature of about 3800~K and there is a large cool spot with a temperature of about 3400~K that can be found around latitude 20--60$^\circ$ and longitude 220--280$^\circ$. The magnetic field within this spot reaches about 0.5~kG for the Stokes $IQUV$ map and about 0.3~kG for the Stokes $IV$ map. The strongest feature, the meridional spot with a strength of 2.1~kG, has a temperature of about 4000~K. The large cool spot has a trailing, smaller spot with a temperature of about 4000~K, which seems to coincide with the strongest negative magnetic feature in the meridional field component map.

\section{Discussion}
\label{dis}

As part of the self-consistent iterative ZDI procedure, we started mapping the surface temperature distribution assuming a null magnetic field instead of doing a magnetic inversion assuming a homogeneous temperature as a first step. Even though most spectral lines are affected by a magnetic field to some extent, this effect is smaller compared to the effect of temperature inhomogeneities on the polarization profiles when dealing with the magnetic field strengths typical of a cool active star. The change in temperature distribution between the consecutive temperature inversions was also very small, even though the fixed magnetic field distribution was significantly different, at least for the two first inversions. The differences between the final temperature distributions of the Stokes $IV$ and Stokes $IQUV$ inversions (Figs.~\ref{rec12}, \ref{rec13}) are small for both observational epochs, even though the corresponding magnetic maps are very different. This allows us to conclude that temperature inversions are weakly sensitive to magnetic fields of this strength and that temperature DI mapping carried out with non-magnetic inversion codes is not significantly biased.

All LSD line Stokes profiles from the two observational epochs show distinct polarization signatures of similar amplitude, suggesting that the activity level of II~Peg remained roughly constant. They do, however, clearly show that the field topology has evolved between the two sets of observations. Some observations from the 2013.05 set overlap in phase with the observations in 2012.75. The Stokes $QUV$ profiles at these phases do not resemble each other; compare for instance profiles around phases 0.15, 0.45, 0.71 and 0.86. As already mentioned in Section \ref{aps}, the mean longitudinal magnetic field values at these phases do not agree as well. Naturally, this difference is also reflected in the corresponding reconstructed magnetic and temperature maps. This suggests that the evolutionary timescale of the surface magnetic field is shorter than eight months. The previous Stokes $IV$ magnetic field study of II~Peg \citep{Kochukhov2013} found evolutionary changes in the radial and meridional magnetic field components within three months. On the other hand, the rms values of the three magnetic components and the field modulus are larger for the 2013.05 set compared to the corresponding values for the 2012.75 set, indicating that the activity level of II~Peg might be increasing. This may be part of the long-term trend identified by \citet{Kochukhov2013}.

The temperature maps from the two sets of observations do show some similarities. Both exhibit a polar spot, with a temperature of 3500--4000~K. There is also another persistent large cool spot, with a temperature of about 3400~K, at approximately the same latitude in both maps. The longitude, however, is slightly shifted between the two sets. Assuming this to be the same spot, this suggests a migration, possibly due to differential rotation.

There is an obvious difference between the magnetic field maps reconstructed using only the Stokes $IV$ spectra and all four Stokes parameters. This is evident from comparing the maps directly and from comparing the Stokes $QU$ model profiles predicted by the Stokes $IV$ inversions to those obtained by the Stokes $IQUV$ inversions. Most surface magnetic features remain hidden when linear polarization is neglected in the ZDI modeling and some features are even recovered with the incorrect polarity. When only Stokes $IV$ data are modeled, 4--6\% of the total energy is distributed between modes with $l$ from 11 to 20 and 70--84\% ends up in modes with $l$ from 1 to 5 (see Figs.~\ref{sep12_mode} and~\ref{sum13_mode}). At the same time, when all four Stokes parameters are taken into account about 20--23\% of the total mode energy is found in modes with $l \geq 11$ and only 33--36 \% is deposited in $l$-modes 1--5. We observe a systematic shift in the spatial magnetic energy power spectrum from a strong dominance of the first few $l$-modes for the Stokes $IV$ inversions to a much broader distribution for the Stokes $IQUV$ inversions. The Stokes $IV$ results cannot correctly reproduce even the lowest order $l$-modes since an inclusion of the Stokes $QU$ data significantly reduces their energies.

The agreement between the model profiles and the observed LSD Stokes $V$ profiles is equally good, independent of whether the Stokes $QU$ data are simultaneously modeled or not. The same set of Stokes $IV$ profiles can hence be fitted by very different magnetic field configurations, implying that the Stokes $IV$ inversions do not provide a unique solution and that Stokes $V$ is insensitive to higher order $l$-modes. When only Stokes $IV$ spectra are used to reconstruct a complex magnetic field, they can, at best, provide an overall, smoothed picture of the magnetic field topology but miss much of the field complexity. Many previous ZDI studies of active cool stars probably suffered from this problem since the reconstructed fields appeared complex even from the Stokes $V$ data alone. This situation calls for inclusion of linear polarization in the ZDI modeling. Unfortunately, the weakness of typical magnetic fields usually prohibits detecting the Stokes $QU$ signatures for all but the brightest active stars. 

Another sign of the shortcomings of Stokes $IV$ modeling versus four Stokes parameter inversions can be found by investigating the changes in the radial, meridional and azimuthal field components. The largest increase in rms value from the Stokes $IV$ to the Stokes $IQUV$ inversion was found for the meridional component in both observational epochs. The large increase in the meridional field rms values in the four Stokes parameter inversions suggests that the meridional component benefits most from the full Stokes vector modeling as discussed in Section~\ref{intro}. However, all three components become substantially more complex and exhibit significant structural changes when we choose to model the Stokes $IQUV$ data with our ZDI code. 

The extended 3D magnetic field topology found for both observing epochs shows more open field lines and a stronger radial field at the Alfv\'en radius in the Stokes $IV$ case compared to the Stokes $IQUV$ case. This is perhaps not too surprising since high order $l$-modes decay faster than low order ones. As discussed above, the magnetic fields reconstructed from the Stokes $IV$ spectra have more energy in the few lowest $l$-modes compared to the Stokes $IQUV$ inversions, meaning that the field strength decreases slower as we extrapolate towards the source surface. In general, the Stokes $IQUV$ inversions imply a more compact stellar magnetosphere (the volume defined by the presence of closed field lines) than would be inferred from Stokes $IV$ inversions.

\section{Conclusions}
\label{con}

Several theoretical ZDI studies, as well as observational studies of magnetic Ap stars, have suggested that four Stokes parameter modeling is highly preferable to Stokes $IV$ reconstructions and generally represents the only way to retrieve a complete and unbiased picture of stellar magnetic field topologies. We therefore initiated systematic, time-resolved four Stokes parameter observations of RS~CVn binaries with the ESPaDOnS spectropolarimeter at CFHT. One of these targets, II~Peg, showed extraordinarily strong linear polarization signatures suitable for detailed ZDI modeling. Here we have presented magnetic mapping of this star based on these four Stokes parameter observations, which is done for the first time for a cool active star. We have combined temperature reconstruction using individual lines with a new Zeeman Doppler imaging method employing LSD Stokes profiles. Using this approach, we were able to obtain successful fits to four Stokes parameter observations of II~Peg at two epochs. This enabled us to perform a detailed comparison between results of traditional Stokes $IV$ inversions and our more sophisticated four Stokes parameter modeling. The main conclusions of our study are the following:
\begin{itemize}

\item
The new approach to magnetic inversions using LSD profiles does not invoke any assumptions about their behavior with temperature or magnetic field. LSD profiles are only used to compress information and the same line-mask is used to derive both the observed and the synthetic LSD profiles. 
\item
The observed LSD line profiles of II~Peg from both observational epochs exhibit clear distortions in Stokes $I$ and clear polarization signatures in LSD Stokes $VQU$ with similar amplitudes. Line profiles from the two epochs overlapping in phase are generally dissimilar, neither do they correspond to the same mean longitudinal field. This difference, also reflected in the reconstructed magnetic maps, implies that the surface field distribution of II~Peg has evolved on a time scale shorter than eight months.      
\item
The temperature maps from the two observational epochs show some common features. The visible rotational pole is cool in both cases. There is also a prominent cool spot at the same latitude, but slightly shifted in longitude.
\item
The magnetic field is predominantly poloidal (56.7--67.5\% of the total field energy) and axisymmetric (56.2--64.6\% of the total field energy) in all magnetic inversions. 
\item
The radial, meridional and azimuthal field components become stronger when four Stokes parameters are used in the ZDI inversion. Their rms values increase and some individual surface features become stronger. The total magnetic field energy increases by a factor of 2.1--3.5. 
\item
The field topology becomes much more complex when the Stokes $IQUV$ data are modeled by ZDI. In terms of the spherical harmonic expansion this means that a larger proportion of the field energy is distributed among higher order modes. 
\item
Stokes $V$ can be fitted equally well by very different magnetic field topologies. Thus, the Stokes $IV$ inversions do not yield a unique solution and are insensitive to higher order harmonic terms. In the case of II~Peg ZDI inversions without the $QU$ data are also not particularly successful in recovering the low-order harmonic components.  
\item
The meridional field component seems to benefit most from the four Stokes parameter modeling, as predicted by theoretical ZDI studies.
\item
Including linear polarization in the magnetic reconstruction process brings out important, and otherwise hidden small-scale magnetic features and reduces crosstalk between the field components.
\item
The extended field topology obtained with the help of potential field extrapolation from the ZDI results is noticeably affected by the difference between the radial field component recovered in the Stokes $IV$ and $IQUV$ inversions. The stellar magnetosphere is more compact in the latter case and the total field energy is lower by a factor of 2.5--3.1 at the source surface assumed to be located at $3R_\star$.
\end{itemize}

\acknowledgements
OK is a Royal Swedish Academy of Sciences Research Fellow, supported by the grants from Knut and Alice Wallenberg Foundation and Swedish Research Council. GAW is supported by a Discovery Grant from the Natural Science and Engineering Research Council of Canada (NSERC). The computations presented in this paper were performed on resources provided by SNIC through Uppsala Multidisciplinary Center for Advanced Computational Science (UPPMAX) under project snic2013-11-24.

\bibliographystyle{apj}
\bibliography{astro_ref_v1}

\end{document}